\documentclass[onecolumn,authoryear]{els-mrw} 

\usepackage{amsmath,amssymb,amsfonts,amsthm,makeidx,graphicx}
\usepackage{txfonts}
\usepackage{helvet}

\usepackage{gensymb} 
\usepackage{xfrac} 
\usepackage{xspace} 
\usepackage{aas_macros} 

\def\HA{\mbox{{H}${\alpha}$}\xspace} 
\def\HB{\mbox{{H}${\beta}$}\xspace} 

\newcommand{\spec}[3]{\mbox{#1\,{\sc #2}\,$\lambda$#3}\xspace} 
\newcommand{\cd}{\mbox{${\mathrm{c\,d}}^{-1}$}\xspace} 
\newcommand{\vsini}{\mbox{$v\sin i$}\xspace}
\newcommand{\kms}{\mbox{${\mathrm{km\,s}}^{-1}$}\xspace}
\newcommand{\gcm}{\mbox{$\mathrm{g\,cm}^{-3}$}\xspace}

\newcommand{\rstar}{\mbox{$R_{\star}$}\xspace}
\newcommand{\Rstar}{\rstar}   

\newcommand{\vrot}{\mbox{$v_{\mathrm{rot}}$}\xspace}
\newcommand{\vcrit}{\mbox{$v_{\mathrm{crit}}$}\xspace}
\newcommand{\vorb}{\mbox{$v_{\mathrm{orb}}$}\xspace}

\newcommand{\omeang}{\mbox{$\omega$}\xspace}
\newcommand{\omelin}{\mbox{$\Upsilon$}\xspace} 
\newcommand{\omeorb}{\mbox{$W$}\xspace}

\renewcommand{\deg}{\degree\xspace}

\newcommand{\req}{\mbox{$R_{\rm eq}$}\xspace}
\newcommand{\rpole}{\mbox{$R_{\rm pole}$}\xspace}

\begin{document}

\chapter{Classical Be stars}\label{chap1}

\author[1,2]{Thomas Rivinius}%
\author[1,3]{Robert Klement}%


\address[1]{\orgname{European Organisation for Astronomical Research in the Southern Hemisphere (ESO)} \orgaddress{Casilla 19001, Santiago 19, Chile}}
\address[2]{ {\tt triviniu@eso.org} }
\address[3]{ {\tt rklement@gmail.com} }


\maketitle

\begin{glossary}[Glossary]
\term{Emission line star} Star with a spectrum showing emission lines, most typically of the hydrogen Balmer series, and potentially other ion species. Only used for emission not associated with an angularly extended nebular region.

\term{Main sequence} Range of photospheric observables into which the vast majority of all stars fall. In physical terms the group of stars that produce energy through core hydrogen fusion. In the spectral range of OBA stars taken to include luminosity classes V, IV, and III. 

\term{Optical range} No generally agreed definition exists, but depending on author it can mean the same as visible range, be widened to encompass the range accessible through the same technological concepts from the ground, i.e., optical telescopes and detector technology, which may be as wide as 330\,nm to 1100\,nm, and some definitions see all electromagnetic radiation from 100\,nm to 1\,mm as optical. In the context of this work meant as 330 to 1100\,nm.

\term{Observables} Properties that can be directly measured from the radiation, waves, or particles received from an object, i.e., without the need for modeling.

\term{Paschen continuum} Featureless, continuous spectrum in the wavelength range between the Balmer and Paschen jumps at 346.6 and 820.4\,nm, respectively.

\term{Photosphere} The region of a star from which the observable optical light originates, usually considered to be the surface.

\term{Shell star} Star with a spectrum showing sharp absorption (shell) lines formed under conditions cooler than and not typically associated with the spectral type.

\term{Superionization} Presence of ionized species with ionization temperatures above the local thermal equilibrium.

\term{Visual range} Range of the electromagnetic spectrum accessible to the human eye, about 380 to 700\,nm. 

\end{glossary}

\begin{glossary}[Nomenclature]
\begin{tabular}{@{}lp{34pc}@{}}
Ae star& A-type emission line star\\
AM & Angular momentum\\
Be star& B-type emission line star, commonly understood to mean classical Be stars only\\
BeXRB & Be X-Ray binary \\
CBe star& Used by some authors to explicitly specify a classical Be star\\
\cd & Cycle per day. Frequency unit equal to to $86400^{-1}$\,Hz\\
e & Spectral type suffix to describe the presence of emission lines\\
E/C & Peak height of an emission line in units of the local continuum\\
$[$e$]$ & Spectral type suffix for the presence of forbidden emission lines\\
HRD & Hertzsprung Russell Diagram \\
HST & Hubble Space Telescope \\
IR & Infrared \\
LC & Luminosity class\\
LPV & Line profile variability\\
MS & Main sequence \\
MTD & Magnetically torqued disk \\
MW & Milky Way \\
n or nn & Spectral type suffixes to denote broad and shallow absorption lines\\
NIR & Near infrared\\
NLTE & Not in local thermal equilibrium\\ 
NRP& Non-radial pulsation\\
Oe star& O-type emission line star\\
sdOB & Subdwarf OB-type star\\
SED & Spectral energy distribution\\
SPB & Slowly pulsating B star. Previously also known as 53\,Per-type stars\\
SpT & Spectral type\\
sh & Spectral type suffix to denote a shell star\\
UV & Ultraviolet\\
VDD & Viscous decretion disk\\
V/R & Violet-to-red peak height ratio of a double peaked emission line\\
WCD & Wind compressed disk \\
YSO & Young stellar object\\
ZAMS & Zero-age main sequence \\
\end{tabular}
\end{glossary}

\begin{abstract}[Abstract]
Classical Be stars, the "e" standing for the presence of spectroscopic line emission, are main sequence stars of spectral type B that are able to form a gaseous disk in Keplerian motion from star-ejected matter. The main driver of this capability is the rapid surface rotation, which might be acquired via binary interaction or through internal stellar evolution, but additional mechanisms, such as nonradial pulsation, usually enable a star to become a Be star well below the actual critical rotation threshold. The angular momentum loss through the disk then keeps the star below the critical rotation value for the rest of its main sequence life span. It is one of the oldest standing research fields of astronomy, since the first discovery of a Be star in 1866. The article, therefore, not only presents the properties of Be stars, but as well the history of the field. The current main research topics, discussed in greater detail, are: 1) the variability of the central object and the nature of Be stars as nonradially pulsating stars, 2) the physics of the viscosity governed circumstellar disk and its variability, which can serve as proxy for the more common, but typically harder to observe viscuous accretion disks, and 3) the role of binarity both in the formation of Be stars as rapid rotators, and as well their impact on the observed properties of these stars.
\end{abstract}

{\bf Key points:}
\begin{itemize}
\item Classical B-type emission line (Be) stars are the class of the
  most rapidly rotating non-degenerate stars.
\item They are non-radially pulsating stars with great asteroseismic potential, once the theoretical difficulties posed by the rapid rotation are overcome.
\item Most early type Be stars are spun-up by binary evolution. This is less certain for late type Be stars, which might have experience evolutionary spin-up.
\item Their self-ejected gas disks, responsible for the line emission, are in Keplerian rotation, goverend by viscosity, and the means by which the star sheds angular momentum.
\end{itemize}

\section{Introduction}\label{sec:intro}
Stars of spectral type B exist in a temperature range between about 10\,000 and 30\,000\,K. It is quite natural for any gaseous matter in the stellar vicinity, exposed to such a UV intense ionizing radiation field, to produce line emission in the spectrum. Thus, the broadest possible definition for a Be star, namely as B type star showing a spectrum with line emission, lacks sharpness. Such a definition would include interacting binaries, B-type supergiants with strong stellar winds, YSOs such as Herbig AeBe stars, stars with magnetically confined co-rotating clouds, and even B stars surrounded by nebulosity.
With that in mind, a Be star is commonly understood as a \emph{non-supergiant B-type star that has a history of showing hydrogen lines in emission}. Among those, the vast majority are "classical Be stars" \citep[see also][for an in-depth review]{2013A&ARv..21...69R}. For the purpose of rapid classification in survey data this definition of Be stars is sufficient, even if it leads to misclassifications in a small minority of cases. Detailed analysis reveals further meaningful distinctions, but since that relies on high quality or long-term observations, these are more difficult to apply. 

\begin{figure*}[h!]
\begin{center}
\includegraphics[width=0.6\textwidth]{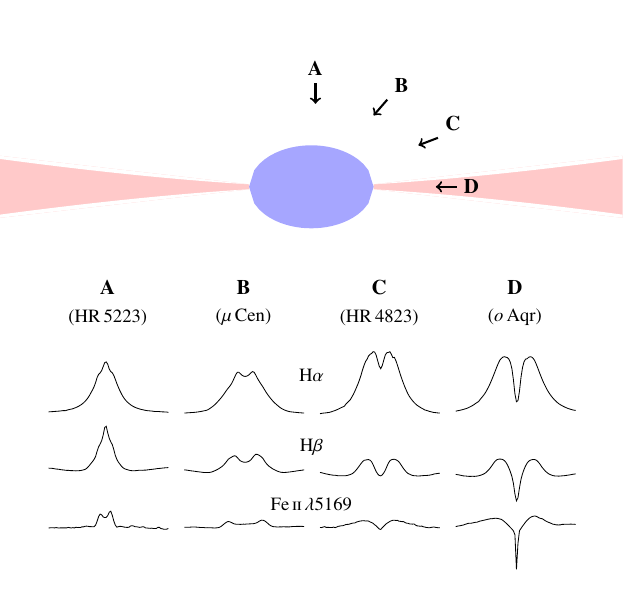}
\end{center}
\caption[xx]{\label{fig:bescheme}Top: Schematic representation of \citeauthor{1931ApJ....73...94S}'s (\citeyear{1931ApJ....73...94S}) idea of a critically rotating B star with an equatorial disk, but showing a more modern understanding of the flaring vertical disk structure. Below: Actual emission line profiles of selected spectral lines of three pure emission line stars (left) and one that is showing as well shell properties (rightmost) and how they differ as a consequence of inclination. From \citet{2013A&ARv..21...69R}. 

%
}

\end{figure*}

Concisely, a classical Be star is a very rapidly rotating B-type star of luminosity class (LC) V to III, surrounded by a gaseous disk that is formed of self-ejected matter, as sketched in Fig.~\ref{fig:bescheme}. The rapid rotation can be the result of binary interaction or internal stellar evolution. Once ejected, part of that matter settles into a disk with Keplerian orbital motion and its further evolution is mainly governed by viscosity. The disk density and structure is dynamic and might be variable to the point where the disk is completely lost and later possibly reformed, although Be stars with disks in a steady state exist as well. In spectroscopic terms, these disks may express themselves both in line emission, but also as sharp absorption lines across the entire spectrum.  In spectral classification, the  appearance is distinguished by the Spectra type (SpT) suffixes "e" or "sh", respectively, and in the latter case a Be star is also dubbed a "shell star". Similarly, optical photometry might show a flux excess for non-shell stars, and a diminished one for shell stars. Whether a Be star presents itself as shell star or not is determined by the inclination angle of the disk, as shown in Fig.~\ref{fig:bescheme}.

\subsection{Structure of the article}

In the following, Sect.~\ref{sec:def} gives a summary of the general physical properties of a Be star. 
The historical path towards that understanding is outlined in
Sect.~\ref{sec:hist}, mentioning the major historical debates in Sect.~\ref{subsec:debates} and ending with the main issues investigated today in Sect.~\ref{subsec:current}.
For a more detailed view, Sect.~\ref{sec:star} discusses the central stellar objects of Be stars, and in particular focuses on the stellar rotation in Sect.~\ref{subsec:rot} and the short-term periodic variability in Sect.~\ref{subsec:puls}.
Section~\ref{sec:disk} describes the circumstellar disk, first for a stable Be star with a steady state disk in Sect.~\ref{subsec:disksteady}, then for a disk including its various potential variabilities in Sect.~\ref{subsec:diskvar}.
Binarity of Be stars is considered in Sect.~\ref{sec:bin}, both in terms of how a part of Be stars can form as binary products in Sect.~\ref{subsec:binprod}, and as components of current binary systems in Sect.~\ref{subsec:curbin}.
Finally, Sect.~\ref{sec:concl} presents conclusions and further reading.


\begin{figure*}[t]
\begin{center}
\includegraphics[width=0.8\textwidth]{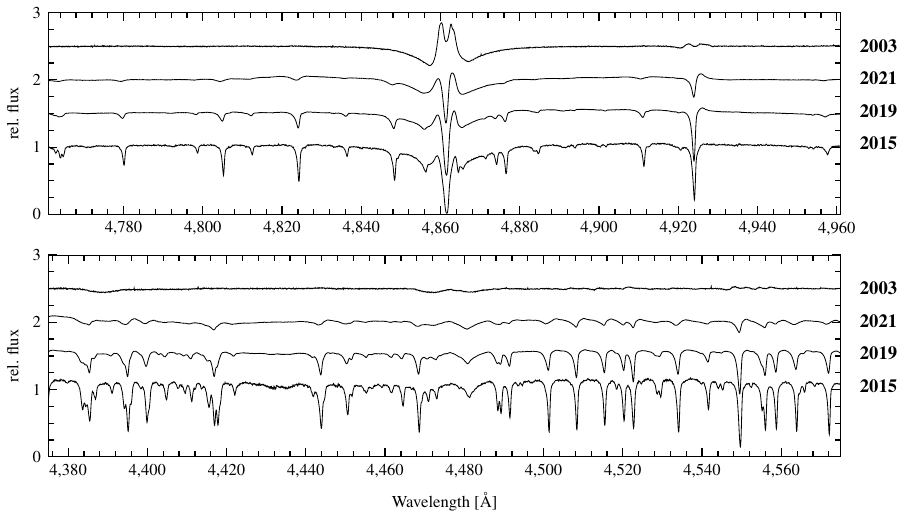}

\end{center}
\caption[xx]{\label{fig:shell}
The late type Be star Pleione (28\,Tau) showing a series of different spectral appearances. Pleione is one of the few Be stars with a non-aligned, precessing disk and oscillates between strong shell (2015) and pure emission (2003) appearance. Upper panel: Region centered on \HB.
Lower panel: Region centered on \spec{He}{i}{4471}/\spec{Mg}{ii}{4481}. 
See Sect.~\ref{subsec:diskvar} for details.
}

\end{figure*}

\begin{figure*}[t]
\begin{center}
\includegraphics[width=0.65\textwidth]{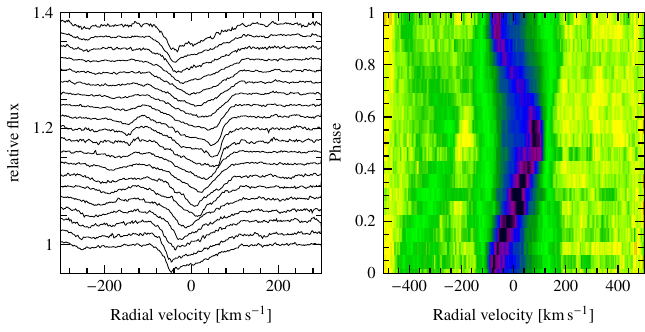}
\end{center}
\caption[xx]{\label{fig:LPV}
Periodic spectral line profile variability of the Be star $\omega$\,CMa, shown for the photospheric \spec{Si}{iii}{4553} line. The weaker \spec{Fe}{ii}{4549} line can be detected bluewards.  The $g$-mode nonradial pulsation expresses itself mostly as a surface velocity field adding to the rotation, leading to a redistribution of the localized line formation to different projected velocities.
A total of 64 spectra, taken in 2000, have been phased with the period of 1.3712\,d into 20 phase bins. Left: Phase binned normalized spectra, vertically offset in steps of 2\% for clarity. Right: Dynamical trail of the phase-binned spectra. Time runs upwards.
See Sect.~\ref{subsec:puls} for details.
}

\end{figure*}

\section{General properties of a classical Be star}\label{sec:def}
The following list introduces observationally determined properties of Be stars as a class. Those most easily measured, like in a simple low-resolution spectrum or through broad-band photometry, are listed first, while the final ones may require detailed modeling of spectrally well resolved interferometric data or long term datasets acquired over decades:
\begin{description}
    \item[Spectral type:] As stated above, line emission emerges easily from any sufficiently dense circumstellar material around a B star. The Be phenomenon is, however, not restricted to B stars. Classical Be stars extend both into the late O and early A spectral types, then classified as Oe and Ae stars, respectively. In earlier O-type stars radiative forces may prevent any ejected material from settling into a Keplerian disk
    \citep{2004AN....325..749N}, 
    while in late A-type stars such disks might exist, but are not creating detectable line emission due to the lack of ionizing photons from the central object. Such stars can be identified as A-type shell stars when the disk is seen edge on, but disks in such later spectral types can also be fossil remnants of star formation, as in $\beta$\,Pic. A distinction between these is very difficult with optical data alone, but easily achieved with infrared data that can reveal the presence of dust  \citep[which is not present in Be stars, but only in fossil disks, see below][]{1996ApJ...471L..49G}. 
    Rapid rotation is a physical property of all Be stars, as laid out in the previous section, but the inclination dependence of the rotation on the spectroscopic observables makes it unsuitable for identification and definition purposes.
    \item[Luminosity class:] The ejection of the disk material from the star, and subsequent settling into a Keplerian disk, requires rapid rotation of the central object, even though the majority of Be stars do not reach critical rotation, i.e., the rotation speed at which the centrifugal forces at the equator become equal to the gravitional ones. The precise values of average Be star rotation and their limiting thresholds are debated. As the stellar evolution beyond LC\,III causes a rapid decrease in critical rotation fraction, Be stars cease to exist beyond that limit.
    \item[Spectral energy distribution:] The UV-SED of a B emission line star does
    not differ significantly from the one of a non-Be star, but the Paschen continuum can be brighter. Shell stars, on the other hand, show highly veiled ultraviolet spectra
    and their Paschen continua are typically dimmer. Depending on the amount of circumstellar matter, the
    spectra become dominated by free-free and free-bound emission from the disk in the near- to mid-infrared region.
    In the far IR, the observed spectral energy distribution index $a$ ($S_\nu \propto \nu^a$) changes from $a = 0.6$ to $a > 1$ in the radio regime, indicating some structural change far away from the central star. Stars showing the photometric long-term changes associated with the formation and decay of the disk, are historically classified as $\gamma$\,Cas variables, or "GCAS" in the General Catalog of Variable Stars (GCVS), but the new variable class "BE" has been introduced in the 5$^\mathrm{th}$ edition of the GCVS to include the various other types of variability seen in Be stars as well. Outside of the context of the GCVS, the moniker "$\gamma$\,Cas analogs" is becoming increasingly common to refer to a specific type of X-ray behaviour, see below
     \citep{2019ApJ...885..147K}.
    \item[Polarized light:] The continuum spectra of Be stars are linearly polarized due to the re-processing of photospheric light in the disk.
    The amount of intrinsic polarization can be as high as 2\% and, as a statistical property of all Be stars, depends on the disk inclination, with  polar and equatorial orientations having lower values due to symmetry and self-absorption, respectively.
    In any given star, though, the polarization degree also scales with
    the strength of the line emission. It may also vary with other quantities, like the V/R ratio, while
    the polarization angles are typically constant and perpendicular to the disk, indicating that disks generally have a stable orientation.
    \item[Emission/Absorption:] The signature observable of a rotating disk is the double peaked emission line profile.     
    The detailed appearance of the disk in the observables depends strongly on the inclination it is viewed. In most cases the disk is aligned with the stellar equator, and the disk inclination is identical to the stellar one. A pole-on star presents a face-on disk, and an equatorially seen star an edge-on disk. Due to the density of the disk, the latter case can result in very strong absorption lines of numerous ions, completely obscuring the stellar photospheric spectral properties, as shown in Fig.~\ref{fig:shell}. The peak separation of the emission, as a statistical property of the class, is correlated with the projected stellar rotational velocity, \vsini. In individual stars, it can vary somewhat with disk size, and hence the average radius of the line formation region, where larger, well developed and as well decaying disks tend to have smaller separation than small, newly forming disks.
    \item[Self formed gaseous disk/Absence of dust/No forbidden emission:] The disks of Be stars consist of gas and plasma showing the photospheric chemical abundance pattern. The disk temperature is between 7000\,K and 20\,000\,K, depending on spectral type and density. A good rule of thumb is to expect line formation corresponding to a disk temperature of about \sfrac{1}{2} to \sfrac{2}{3} of the stellar effective temperature.
    This is too hot to form dust, so that the presence of dust is usually a sign for an accretion disk or a fossil disk in a young object, where the dust may have existed as part of a molecular cloud already before a disk was formed.
    Further evidence for the origin of the disk not being fossil, but self formed, lies in its variability. Be star disks are well known to build-up and decay on the time-scale of years to decades, and in particular for the earlier B subtypes the disk is almost never observed in a steady state, but perpetually undergoing minor outbursts and decay events.
    A confusion with steady-state, fossil disks is possible only for the latest B subtypes, and Ae stars, where gas viscosity and stellar radiation properties lead to much longer time-scales of disk variability, so that gas disks could actually be long-term stable around these cooler objects. But even then a distinction is likely possible when looking at dust emission, as explained below.
    From interferometric observations and shell star statistics, the disk is known to be geometrically thin, with a half-opening angle below 20\deg.
    Some B-type emission line stars may also present forbidden line emission. These are classified as B[e] stars, and upon closer inspection consist of various subtypes, including supergiants, YSOs, and interacting binaries. In contrast to Be star disks, B[e] star disks are in a suitable density/temperature range to form dust in their outer regions and show large IR flux excesses.
    \item[Short term periodic variability:] Almost all Be stars observed with sufficient precision exhibit multi-periodic variability with frequencies between 0.5 and 5\,\cd. Photometric amplitudes are between a few tenths and a few tens of millimagnitudes. Often dozens of periods are found, concentrated in a few groups. Only in some stars of the very latest spectral subtypes no periodicity can be found even in space-based photometry. In many early Be stars the periods with the strongest amplitudes can be detected spectroscopically as well, as seen in Fig.~\ref{fig:LPV}. Historically, stars showing this type of behaviour in photometry have been classified as $\lambda$\,Eri variables, but with a better understanding of stellar pulsation this classification has fallen out of favour, and this type of Be star variability is usually understood as similar to that of the Slowly Pulsating B stars, SPBs.
    \item[Keplerian rotation of the disk:] The mass loss rates from non-supergiant B-type stars are too small to cause noticeable line emission if the material was lost from the system immediately, like through a wind. To create the observed emission strengths, the ejected matter must, at least temporarily, accumulate in a reservoir. While magnetically confined and co-rotating clouds can be such reservoirs, no large scale fields have ever been fopund in any Be star, and they could only hold densities that can at most produce modest \HA emission, and no non-Balmer emission at all. Stars with such magnetospheres are not considered Be stars \citep{2013MNRAS.429..398P}. 
    The only option for material to accumulate to sufficient densities (in a big enough volume) is a Keplerian disk. The rotating nature of the disk is most evidently expressed in a double-peaked line profile with a central depression between the violet and redward peaks, as seen in the lower rows of Fig.~\ref{fig:bescheme}. 
    Deviations from circular Keplerian orbits may create one-armed spiral density waves and induce unequal peak heights. Precession of those orbits then gives rise to the so-called violet-to-red (V/R) cycles (Fig.~\ref{fig:VR}). 
    \item[Viscosity governed disk:] Once the material is in Keplerian motion, its further evolution is mostly governed by its viscosity, although the intensity of the stellar radiation may play a noticeable role through ablation for tenuous disks, for the most early spectral types. The state of the disk then becomes a function of the mass injection history and the viscosity alone \citep{2016MNRAS.458.2323K}. 
    The disk physics at this point is very similar to that of accretion disks, except the net mass flow is directed outwards, instead of inwards. For this reason the phrase "decretion disk" is often used to describe a Be star disk. 
\end{description}

\begin{figure*}[t]
\begin{center}
\includegraphics[width=0.5\textwidth]{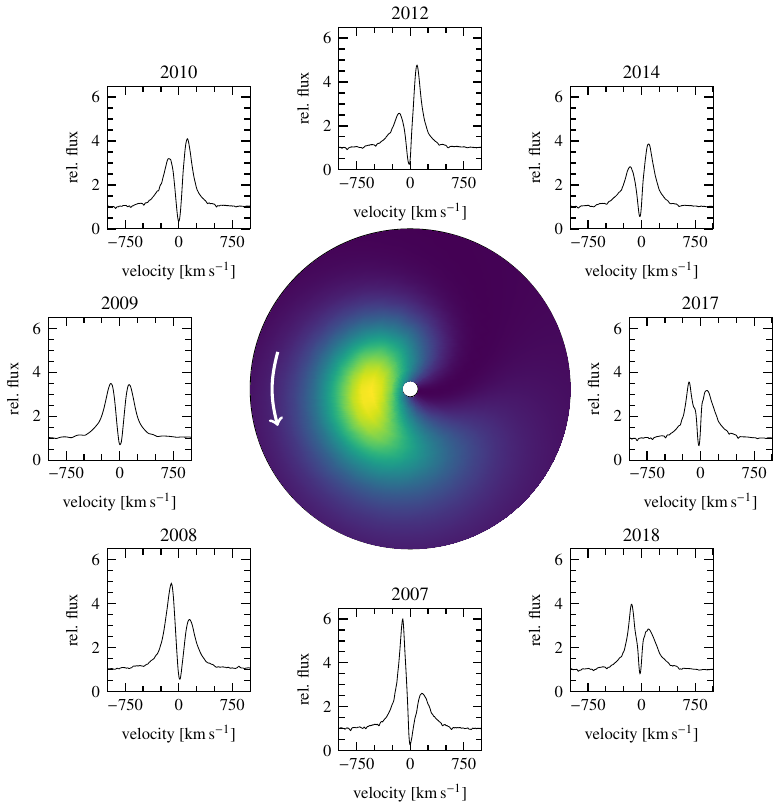}
\end{center}
\caption[xx]{\label{fig:VR}
Cyclic violet-to-red (V/R) peak height variations of the \HA emission line of the shell star 48\,Lib, over about 12 \,years. The schematic picture of the one-armed density wave in the disk in the central panel illustrates how the precession through different viewing angles, indicated by the relative position of the spectral profiles, produces the cyclic variation. The density wave is not in the shape of a radial arm, but of a slightly leading spiral, so that emission lines originating from different radii may show delayed or ahead behaviour with respect to \HA. Disk rotation is counter-clock wise. 
See Sect.~\ref{subsec:diskvar} for details.
%
}

\end{figure*}


\section{History}\label{sec:hist}
The first Be star to be recognized as such was $\gamma$\,Cas, by Father Angelo \citet{1866AN.....68...63S}. While it is not entirely typical for a classical Be star, as will be discussed in Sects.~\ref{subsec:diskvar} and \ref{subsec:curbin},  $\gamma$\,Cas does conform to all the criteria laid out in Sect.~\ref{sec:def}. At the time, spectral classification was still in its infancy, and the paucity of spectral lines in the optical range of OBA stars in general was taken as a sign that these objects would be easier to understand than those of later spectral type. As the taxonomical inhomogeneity of emission line stars began to emerge with more observations over the next several decades, however, the idea that emission line star spectra with only few lines would be easier to explain than the multitude of lines in later spectral types was abandoned. A first breakthrough was achieved by \citet{1931ApJ....73...94S}, who narrowed down the criteria for a Be star by excluding wind-type emission (P\,Cygni) profiles and interacting binaries such a $\beta$\,Lyr. \citeauthor{1931ApJ....73...94S} suggested the emission to come from a circumstellar gas disk around a rapidly rotating B star. This picture, sketched in Fig.~\ref{fig:bescheme}, unified classical emission line and shell stars in a single class and provided a hypothesis on the disk formation mechanism as equatorial mass loss from a critically rotating star. 

The next few decades were largely characterized by efforts to find, catalog, and classify Be stars and their fundamental properties to better define the distinction between classical and other types of Be stars \citep{1979SSRv...23..541S}. 
It was found that classical Be stars represent a considerable fraction of all B type stars, with the current estimates reaching up to about 20\% for the early subtypes. The statistics for later subtypes is less certain due to various detection biases, such as a tendency to have weaker emission in the traditional classification range that only includes H$\beta$, but not \HA, but is at least 10\%, and may even be similar to the earlier type ones. For classical Be stars \citeauthor{1931ApJ....73...94S}'s model remained largely accepted, with the caveat, however, that the stellar rotation was found to be rapid, but not quite critical.

\subsection{Previous debates on the nature of Be stars}\label{subsec:debates}
An era of breakthrough on the physical nature of Be stars began with UV observations from satellite observatories in the 1970s, and continued with the increasing use and improvement of digital detectors. An important early result for hot stars in general was the discovery of "superionization", i.e., the presence of ion species with higher ionization energies than possible in thermal equilibrium around a host star of a given temperature. Since, at the time, this was considered to be a tell-tale sign of a coronal structure, and not as a general NLTE phenomenon in any hot stellar wind, this led to several alternative hypotheses to \citeauthor{1931ApJ....73...94S}'s model for Be stars.
In the coronal wind hypothesis the emission would emerge from a largely spherical coronal region, and V/R variations, such as in Fig.~\ref{fig:VR}, would be the signature of alternating expansion and contraction. 
Another alternative to \citeauthor{1931ApJ....73...94S}'s model was also proposed by \citet{1975BAICz..26...65K}, who argued for Be stars to be a special case of interacting binaries with accretion disks. 
All of these models are discussed in some detail by \citet[][in particular in Part II]{1982bsww.book.....U} and \citet{1988PASP..100..770S}.
None of these hypotheses did prove to be correct, however, when scrutinized with later observing techniques, like X-ray, infrared, and radio observations, as well as interferometry, and \citeauthor{1931ApJ....73...94S}'s model of the circumstellar disk largely still holds.

Another question beyond the original model was cast by the sub-critical rotation of Be stars, namely what else was required to form a disk. 
Gas ejected even from the equator of such a star (i..e., with the largest boost) would lack angular momentum (AM) to settle into a Keplerian disk, and instead fall back to the star. A large number of early type Be stars had been found to show episodic outbursts to be a major factor in fueling the disk, and so mechanisms capable of only providing a steady state mass flow were insufficient. With improving electronic detectors also many early type Be stars were found to exhibit short-term periodic variability of their photospheric line profiles which is unique to Be stars, and this became a main candidate for the process acting in addition to rotation.  Drawing from the statistics of the observed periods, it was suggested that this variability was rotational modulation and possibly related to magnetic fields. The alternative hypothesis to explain the spectroscopic line profile variability (LPV) signature of many early Be stars was non-radial pulsation (NRP). The debate about the nature of the LPV \citep{1994IAUS..162..311B} was largely concluded in favor of NRP when on the one hand multiperiodicity was found to be ubiquitous in Be stars \citep{2022AJ....163..226L}, 
which is difficult to explain with rotation, and on the other hand no magnetic field could be measured in any Be star\footnote{This non-detection rules out large scale fields with a dominant dipolar component at the level of a few Gauss. Strictly speaking, highly structured, locally strong fields remain possible, though none have been found yet, either. The debate, however, was about the nature of the LPV and corresponding photometric variations, and magnetism is firmly ruled out as cause for that.} \citep{2016ASPC..506..207W}. 
If and how NRP is indeed the missing mechanism for AM transfer remains unclear, however.

The third debate concerned the nature of the disk rotation, namely whether it was really Keplerian or rather an AM conserving, outflowing structure. Observational evidence was increasingly collected for the Keplerian hypothesis, first spectroscopically and then, decisively, through spectrally resolved interferometry of the \HA and Br$\gamma$ lines. However, at about the same time a natural disk formation process in the form of a combined wind-rotation scenario was suggested by \citet{1993ApJ...409..429B}. 
This process would always occur in a wind of a rapidly rotating star and was called the wind-compressed disk (WCD). Without further modification a WCD would create an outflowing disk, not a Keplerian one, and so it was adapted as the magnetically torqued disk (MTD) model  by \citet{2002ApJ...578..951C} 
to produce a Keplerian disk. More detailed modeling work, however, showed that the basic WCD mechanism would be inhibited by secondary nonradial, radiative driving effects on the outflow, even leading to a polarly enhanced stellar wind \citep{1996ApJ...472L.115O}.

\subsection{Current drivers of Be star understanding}\label{subsec:current}
With these debates settled, the theoretical work on Be stars went from an era of largely qualitative and simple toy models into the quantitative and physical domain. Most progress has been made on the disk. The dynamical structure of a viscous decretion disk (VDD) can be modeled with both smooth particle hydrodynamics and analytical models \citep{2001PASJ...53..119O}. 
This disk structure can then be used to compute observables using Monte-Carlo radiative transfer codes \citep[e.g.][]{2017IAUS..329..390C}. 

On the observational side, optical and NIR interferometry have settled questions about disk shape and dynamics for good \citep{1997ApJ...479..477Q,2012A&A...538A.110M}, 
and are now used to investigate the binarity of Be stars \citep{2024ApJ...962...70K}. 
Be star binarity is also in the focus of X-ray observations as the main driver of producing high-energy electromagnetic radiation \citep{2011Ap&SS.332....1R}, 
With the relative abundance of bright Be stars, photometric survey satellites such as CoRoT, Kepler, the BRITE constellation, and TESS continue to produce a vast collection of data over long periods of time, ideally suited to further deepen the understanding of NRP and other variability in Be stars \citep{2022AJ....163..226L}. 
Also owing to the relative brightness of these objects even citizen science, in the form of highly dedicated amateur spectroscopy observers, is making valuable contributions to the field.


\section{Central object}\label{sec:star}
The central object in a Be star is an ordinary, although rapidly rotating B-type star. This does not mean every such object will become a Be star. There are many rapidly rotating B stars which are not known for any emission line periods, typically classified as Bn stars. Since a Bn star is recognizable as such only when viewed equatorially, they are equivalent to the group of shell stars, rather than all Be stars. However, they statistically differ from Be stars, as discussed below. 

The stellar wind of a Be star is identical to that of a non-Be star of the same type when seen from polar and intermediate inclinations, but becomes more dense and slowly out-flowing towards the equatorial viewing angle. This is is a consequence of radiative ablation of the outer vertical layers of the disk at intermediate stellar latitudes, and for an equatorial orientation obscuration of the photosphere when viewed through the disk will occur. 
Magnetic fields have never been measured in any Be star, despite dedicated search programs that would have found even weak global magnetic fields of a few Gauss. In contrast, the incidence of such magnetic fields in other O and B-stars is about 10\%. In other words, the phenomena of Be stars and global magnetic fields seem to be mutually exclusive. That newly ejected circumstellar material cannot easily settle into a Keplerian disk in the presence of a magnetic field was also theoretically found, in the evaluation of the MTD hypothesis (see Sect.~\ref{subsec:debates}).

The mechanism of matter ejection that replenishes the disk does not need to be directional or even equatorial, though. An isotropic outburst, if sufficiently large in magnitude, would naturally produce particles with suitable parameters to settle into a Keplerian disk, and the rest would return to the star or escape the system quickly. Both episodic and continuous ejection activity outside the equatorial plane would produce particle orbits over the entire range of orbital nodes. Thus initially inclined particle orbits would circularize in the equatorial plane, driven by viscous interaction, either with a pre-existing disk or with material ejected in other, overlapping, outburst events. There are many examples of such outbursts being episodic, since that is most obvious in the observables, but that does not exclude continuous feeding of the disk, neither the possiblity of outbursts being so frequent that they appear continuous. Observations suggest the latter happens in some stars to at least some degree.

\begin{figure*}[t]
\begin{center}
\includegraphics[width=0.88\textwidth]{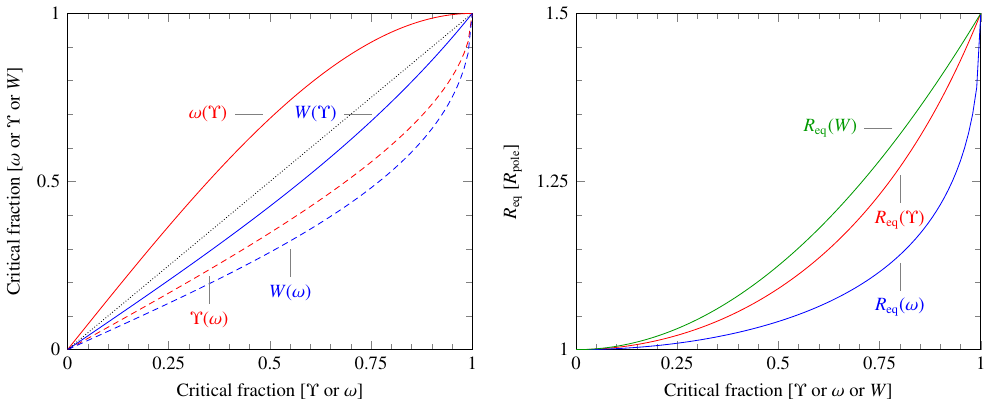}

\end{center}
\caption[xx]{\label{fig:critrot} Relations between the different definitions of critical fraction and oblateness of a star as a function of critical fraction.
Left: Conversions of \omelin, \omeang, and \omeorb as functions of
\omeang and \omelin. The dotted line marks the identity relation.
Right: \req as a function of \omeang, \omelin and \omeorb.
From \citet{2013A&ARv..21...69R}. 
%
%
%
}

\end{figure*}

\subsection{Rotation}\label{subsec:rot}
When dealing with rapid stellar rotation, the typically used parameter is the so-called critical fraction. It is normally calculated as the ratio of the actual to the critical rotation of a given star, either in linear or angular velocity:
\[
\omelin = \frac{\vrot}{\vcrit} = \vrot \sqrt{\frac{3}{2} \frac{\rpole}{GM_\star} }
\hspace{1cm}\mathrm{and}\hspace{1cm}
\omeang =  \frac{\Omega_\mathrm{rot}}{\Omega_\mathrm{crit}} = \Omega_\mathrm{rot} 
\sqrt{\frac{27}{8}\frac{R_{\rm pole}^3}{GM_\star}.}
\]
Since the equatorial radius expands toward $1.5R_\mathrm{pole}$ when approaching critical rotation (see the shape of the star in Fig.~\ref{fig:bescheme}), the critical fraction does not indicate the actually required boost to launch photospheric material into orbit for a subcritically rotating star.  For the problem of Be star disk formation, however, that boost, both in terms of velocity and angular momentum, is the relevant parameter.
For this reason, it is useful to deviate from the common definitions for critical rotation and to define Be star rotation in terms of the Keplerian orbital velocity at the current equatorial stellar radius:
\[
\omeorb = \frac{\vrot}{\vorb} = \vrot \sqrt{ \frac{\req}{GM_\star}. }
\]
The relations between \omelin, \omeang, \omeorb, and \req are shown in Fig.~\ref{fig:critrot}.
Rapid stellar rotation when $\omeorb$ approaches unity introduces complications to the determination of the fundamental stellar parameters.  For a non-rotating star with MW-typical metallicity two independent parameters, like effective temperature and surface gravity, for example, are sufficient to constrain the spectrum of the object, and thus its fundamental parameters. For a rapidly rotating star four independent parameters are needed for the star\footnote{
As an example for an independent set one can take polar radius, bolometric luminosity, mass, and critical rotation fraction. However, the choice varies between authors, as these parameters can be transformed into others, such as replacing luminosity and rotation fraction with polar temperature and oblateness, respectively.}, and an additional one for the inclination under which it is observed. 

Further problems come from rotation induced effects on the stellar photosphere, mostly that the equatorial regions of a rapidly rotating star are cooler and less luminous than the polar ones. This is called "gravity darkening", or "von Zeipel effect", and causes the most rapidly rotating parts of the stellar photosphere to be the ones contributing least to the observables. Because of that effect, rapid rotation above $\omeorb \gtrapprox 0.8$ may not be well distinguishable by the usual means of spectroscopy, but while this may introduce a bias, other ways to estimate the rotation exist, and the hypothesis of critical rotation can be excluded as a general property of Be stars, though it might be realized in a few of them.

With all the above difficulties in mind, observationally it is found that Be star rotation statistics depends on spectral subtype. There is a threshold of $\omeorb\gtrapprox 0.7$ to become a Be star of any subtype, but Be and non-Be stars coexist above that value, until another threshold is reached, above which only Be stars exist. This latter one increases from early subtypes, where it is only a little above the first one, to late subtypes, where it is more like  $\omeorb\gtrapprox 0.95$. In agreement with that, Bn stars are more likely to be of late than of early  sub-type.

Stars can acquire rapid surface rotation in different ways. The original rotation at birth is well known to have too few rapid rotators to explain all Be stars without further spin-up mechanism. The stellar evolution over the MS has the effect of increasing $\omeorb$, because as the core slowly contracts and speeds up rotationally, the photosphere expands, causing $v_\mathrm{crit}$ to decline. At the same time, the mismatch between the core and envelope rotation is being lowered through AM transfer between the two regions, i.e., AM is transported from the core to the surface. This is called evolutionary spin-up, but it must be stressed that firstly, this is the case in terms of $\omeorb$, but not necessarily $v_\mathrm{rot}$ and secondly, the efficacy of the process may vary with spectral type and become negligible for the earliest B subtypes. For a star that started its main sequence lifetime not too far below the critical rotation threshold, this alone might be sufficient to turn it into a Be star, and to force it to remain one for the rest of its MS lifetime. 

However, stellar evolution is not the only mechanism that can push the rotation of a B star into Be star territory. The alternative to acquire rapid rotation is due to binary interaction, where the Be star has a history of accreting mass and AM from an evolving companion, as will be discussed in Sect.~\ref{subsec:binprod}. Both mechanisms produce more Be stars close to the end of the MS than on the ZAMS, which is in agreement with the observed distribution of Be stars in the HRD.

\begin{figure*}[t]
\begin{center}
\includegraphics[width=0.8\textwidth]{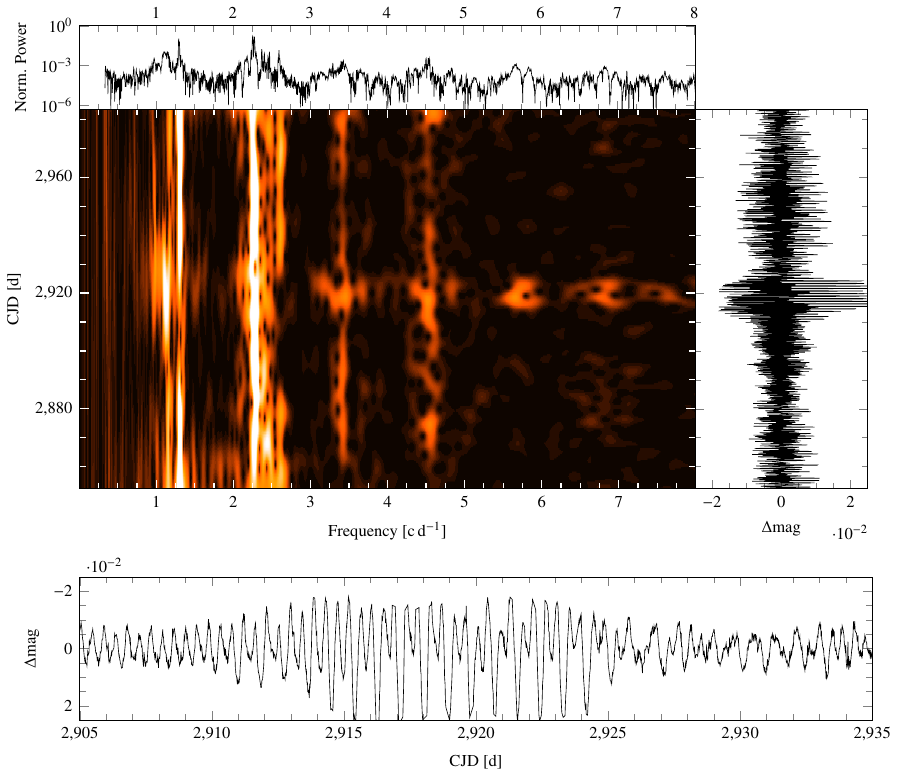}

\end{center}
\caption[xx]{\label{fig:pulsE6}Overview and analysis of the short-term photometric variability of the low inclination B3{\sc IV}e  star CoRoT\,102686433. The outburst seen here was accompanied by a small increase in brightness of about ten millimagnitudes, which was removed for clarity.
The light curve is shown to the right, the overall Fourier transform to the top, and the wavelet analysis in the center. Time runs upwards. In the middle of the observations a clear outburst happened, marked by the strongly enhanced variation amplitude and appearance of a strong additional frequency at about 1.2\,\cd and broad activity over almost the entire frequency band. The lowermost panel shows the time around the outburst in detail. The relative amplitudes of the sharp and well-defined photospheric frequency peaks change a few weeks before the outburst, and return to their original values afterwards.
}

\end{figure*}

\subsection{Nonradial Pulsation}\label{subsec:puls}
Short term multi-periodic variability up to about 5\,\cd is found in almost all Be stars. The CoRoT satellite photometry mission, for instance, observed 31 Be stars, out of which only two, both in the exoplanet field and of SpT B9, were not variable at the level of above 0.1\,mmag. Three more stars of such late type were observed in a more sensitive setup in the seismology field, and found to be variable with amplitudes larger than 0.01\,mmag. This suggests that non-detections are due to instrumental sensitivity, not absence of the phenomenon. While the frequencies are typically long-term coherent, their amplitudes can vary strongly, even on short time scales. 
The multitude of frequencies may be subject to non-linear interaction and energy re-distribution between the modes in the interior of the star, contributing to the changing amplitudes. Unfortunately, the rapid rotation adds severe complications to the asteroseismic interpretation of Be stars, that are not yet overcome. Once this has been achieved, however, asteroseismology will undoubtedly provide answers to the remaining questions of the outburst mechanism, but as well on the stellar interior rotation and structure.

The variations are multiperiodic, and for early subtypes the periods are  concentrated in a few regularly spaced groups. Amplitudes grow towards earlier subtypes, reaching a maximum around B1.5, and for those often one or a few of these periods manifest as spectrscopically detectable LPV. When modeled, most of this  LPV, such as shown in Fig.~\ref{fig:LPV}, is well reproduced by nonradial, retrograde  sectorial $\ell=2, m=+2$ pulsational modes. Typical model amplitudes for the strongest mode are about 15\,\kms, but in the extreme case of $\omega$\,CMa, shown in Fig.~\ref{fig:LPV}, a pulsational velocity amplitude of 35\,\kms was needed to reproduce the observed LPV.
In some stars also other modes are detected, sometimes alone and sometimes in addition to a main retrograde one. In a few stars, the combination pattern of these modes have been observed to be connected to the ejection of material feeding the disk when the combined amplitude is high, suggesting that for the disk formation in at least some objects NRP is the additional, contributing factor beyond rotation. 

A typical example of the photometric variability of an early subtype Be star as seen from space is shown in Fig.~\ref{fig:pulsE6}. The description of this variability can be applied to almost any outbursting early-type Be star:
There are a large number of persistently coherent, but amplitude variable modes that originate in the photosphere. Whether the amplitude variation is purely due to superficial beating phenomena or points to an variable energetic coupling between the various modes inside the star is unknown at this point. In general, there typically is a strong frequency group around 0.5 to 2\,\cd. 

Often, further observed frequencies can be expressed as either differences or sums of frequencies from this fundamental group, although the precise physical mechanism creating variability at such frequencies is unclear, since a purely linear superposition would not do so. As a result, many higher frequencies are at approximate multiples of the frequencies in this fundamental group, giving rise to complex shape and amplitude variations between individual maxima and minima in the light curve, as seen in the lower panel of Fig.~\ref{fig:pulsE6}:
The light curve changes from a fairly regular pattern, where a frequency of about 2.25\,\cd dominates just before the outburst, to a typical "double-wave" pattern during the outburst. The strong frequency at about 1.1\,\cd, present only in outburst, might be caused by the orbital motion of a cloud of newly ejected material, before it circularizes into a ring. Additional variability of circumstellar origin appears during an outburst also at higher frequencies in a very broad, almost white noise-like pattern. 
Post outburst, the variability recovers into irregular looking changes. However, the apparent regularity before the outburst is an effect of the amplitude of other frequencies decreasing, and only a single one prevailing, while the light curve after the outburst is in fact not truly irregular, but rather the complex combination pattern of all those frequencies returning to strength. 

Sharply defined, i.e., long term coherent photospheric frequencies are not exclusive to Be stars, and one finds other non-radially pulsating objects, such as $\beta$\,Cep stars among early subtypes, and "slowly pulsating B stars", or SPBs, among the later subtypes. The latter can look very similar to Be stars, as their variations occupy a similar frequency range, and in particular in the later Be subtypes, where such outbursts are not as often seen as in the earlier subtypes, isolated frequencies, rather than groups, are more common. The main difference between SPBs and Be stars as groups are the rotational velocities, which are generally slower in SPBs, and the cutoff of SPBs towards earlier subtypes. With that in mind, there is some overlap, and stars exist that are simultaneously classified as Be and either $\beta$\,Cep or SPB stars, including the one shown in Fig.~\ref{fig:pulsE6}. Since different pulsation types probe different regions of the stellar interior, such stars are among the most promising for asteroseismology.

In summary, the broad bumps in the power spectrum during outburst are  exclusive to Be stars, as they are connected to the ejective mass loss and subsequent circularization. That amplitudes of the various modes can vary rapidly also in non-Be stars, but there this is a more regular phenomenon and possibly connected to beating effects. The irregularity of Be star pulsation modes dis- and re-appearing is observationally connected to outbursts, but to what degree this is a physical effect of the pulsation itself and to what degree veiling effects by the ejected material are responsible, while the actual pulsation mode remains unchanged, is not clear. Otherwise Be stars seem to fit the idea of an extension of well known pulsating variable types towards more rapid rotation. Although the pulsational behaviour seems often correlated to the outburst activity in one way or the other, the current theoretical understanding of pulsation does not provide a mechanism by which these outbursts could be of sufficient magnitude to create a disk. This would be straightforward only for values of $W\gtrapprox0.95$, where a single strong pulsation mode could bridge the gap between \vrot and \vorb and cause a disk to form. This is a threshold which the majority of Be stars  most likely do not reach, but neither do the majority of Be stars only have a single pulsation mode. 
If NRP is responsible, which seems the most likely judging from observational evidence, a physical explanation will probably have to wait for an understanding of non-linear interactions between multiple pulsation modes, both on the surface and in the stellar interior.


\section{Disk}\label{sec:disk}%

A Be star acquires rapid rotation either during its main-sequence evolution (Sect.~\ref{subsec:rot}) or through binary interactions (Sect.~\ref{subsec:binprod}). In the first case, AM will continue to be transported to the surface and pushes the star towards critical rotation, in the latter the star might become a critical rotator during the interaction. In both cases the formation of Be disks can be understood as the most efficient means for the rapidly rotating central star to shed this excess of surface angular momentum.
Once the material has left the star, and settled into a Keplerian disk, it retains no "memory" of those processes, i.e., for the further understanding of the disk, the details of its creation mechanism are irrelevant. What distinguishes them from other astrophysical disks are, therefore, the inner and outer boundary conditions, not the physical processes.

\begin{figure*}[t]
\begin{center}
\includegraphics[width=0.8\textwidth]{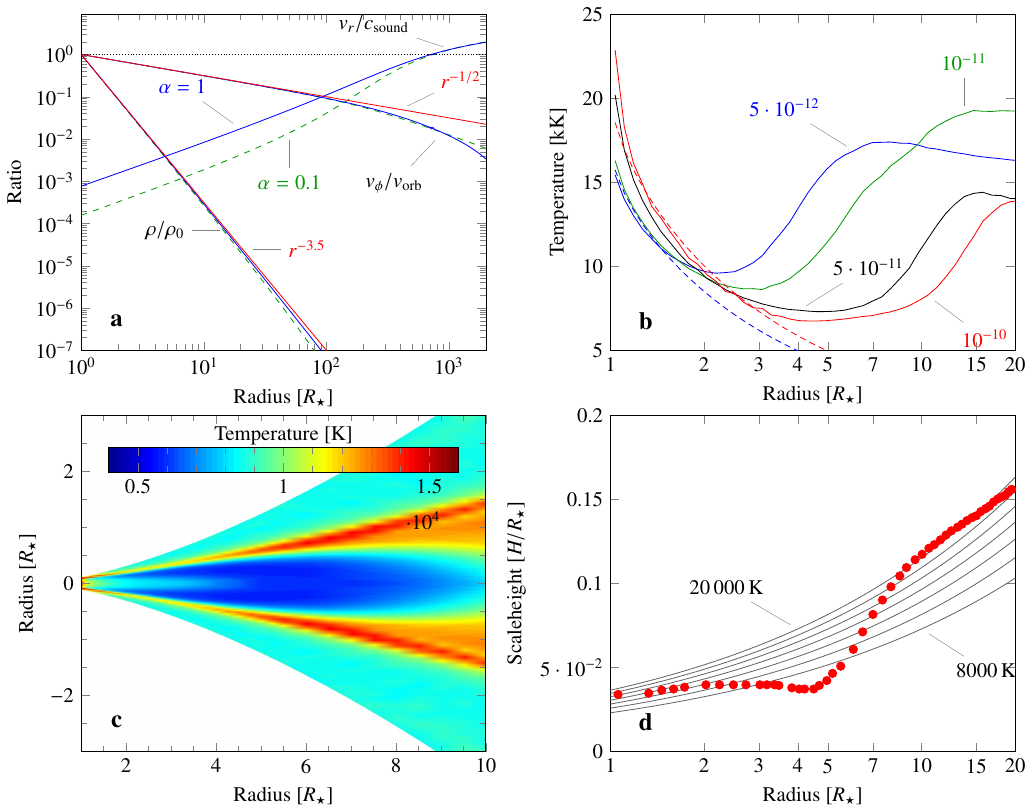}

\end{center}
\caption[xx]{\label{fig:VDD}%
Structure of a steady-state VDD.
{Upper left:} Calculation of the disk density $\rho$, normalized to the value at the base of the disk, radial expansion velocity,
$v_r$, in units of the local sound speed $c_{\rm s}$ and azimuthal velocity, $v_\phi$, in units of the orbital speed at the base of the star, $\vorb$. The solid blue lines are for a viscosity of $\alpha = 1$ and the dashed lines are for $\alpha = 0.1$.
      {Upper right}: Electron temperature along the disk midplane computed with the Monte-Carlo radiative transfer code HDUST. Curves
      are labeled according to their base density $\rho_0$ in \gcm. The dashed
      lines represent a fit of the temperature fall-off. 
      {Lower Left}: Temperature map of the disk (vertical cut) with a base
      density of $\rho_0=1\times10^{-10}$\,\gcm, computed with the Monte-Carlo radiative transfer code BEDISK.
      {Lower Right}: Calculation of the
      scale height of a non-isothermal hydrostatic VDD, compared with the
      corresponding isothermal values in the temperature range of 8000\,K to
      20\,000\,K.  
From \citet{2013A&ARv..21...69R}. 
%
}

\end{figure*}

The VDD model can currently explain the vast majority of observed features, such as the disk kinematics being dominated by Keplerian rotation with only slow diffusion outwards, geometrical thinness, as well as specific observables such as line and (polarized) continuum emission and variability thereof. In this model, turbulent (eddy) viscosity is the mechanism that makes the disks grow radially and that transports the angular momentum from the inner to the outer boundary of the viscous disk. This is the viscosity prescription that was first used in the framework of accretion disks forming around black holes in massive binary systems, and in fact VDDs follow the same physics as these accretion disks save for the inner boundary being a source of material and angular momentum, instead of a sink.

\subsection{Steady state of a Be disk}\label{subsec:disksteady}
A disk governed by viscosity cannot be static. The viscosity will always drive radial flows of material and AM through the disk. A disk can, however, be in a steady state: when the injection of mass and AM at the base is at a constant rate, the disk will  asymptotically achieve a stable structure, where the innermost parts of the disk approach dynamic equilibrium faster than the outer parts. 

In the vertical direction, the disk is in hydrostatic equilibrium. This means that the scale height of the disk, also as a function of radius, is controlled only by the gas pressure and the gravity of the star as
\[
H(r) =  \frac{c_s}{\vorb}  
           \frac{r^{3/2}}{\rstar^{1/2}},
\hspace{1ex}\mathrm{where\ the\ sound\ speed\ }
c_\mathrm{s}=\sqrt{\frac{kT}{\mu m_{\rm H}}}
\]
with $\mu$ the mean molecular weight of the gas, $T$ the electron temperature, and $m_{\rm H}$ the hydrogen mass. In other words, the scale height is proportional to the ratio between the sound speed and the orbital velocity. This is called a flaring disk because the opening angle grows with distance from the star. For a typical Be disk at about \sfrac{1}{2} to \sfrac{2}{3} of the photospheric temperature, the scale height starts with a value of $0.04\,\Rstar$, which corresponds to a half-opening angle of about 2\deg, and
puffs up to $3.5\,\Rstar$ (10\deg) at a distance of 20\,\Rstar from the star.

The viscosity $\nu$ is parametrized by 
\[
\nu = \alpha c_{\rm s} H
\]
where $\alpha$ is a dimensionless parameter that relates the characteristic size of turbulence cells to the disk scale height $H$.
The exact mechanism behind the viscosity is unknown, but it might be connected to the so-called magneto-rotational instability.

Figure~\ref{fig:VDD} summarizes the velocity, density and temperature structure of such a steady state VDD. The density drops with $r^{-3.5}$, where the midplane surface density goes like $r^{-2}$\footnote{This exponent is simply due to the conservation of mass in a two dimensional steady state flow.} and the additional decline is due to the vertical flaring (subpanel d). The outflow velocity increases outwards, as the density declines and the viscuous coupling gets weaker. The disk rotation remains Keplerian until the speed of sound is comparable to the orbital and outflow velocities.  At this point the disk becomes a momentum conserving, outflowing structure, and the outflow velocity is no longer governed by viscosity (subpanel a).

\begin{figure*}[t]
\begin{center}
\includegraphics[width=0.8\textwidth]{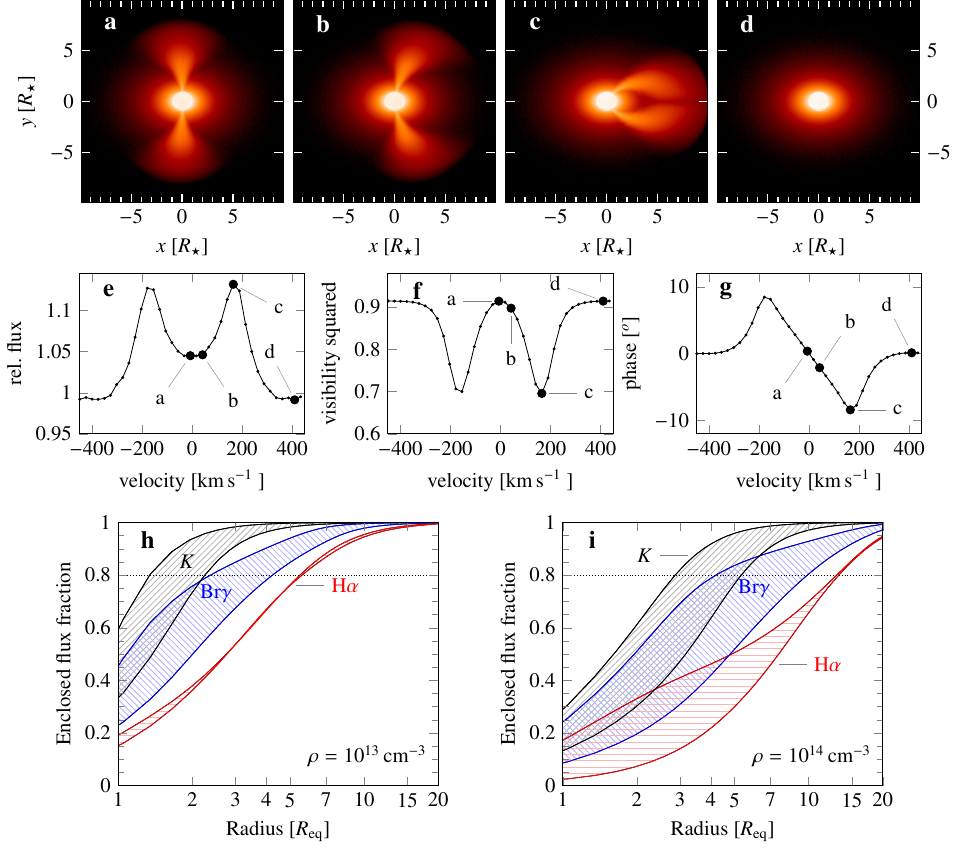}

\end{center}
\caption[xx]{\label{fig:lineform}
{Upper row:} The formation regions of an emission line (Br$\gamma$, panels a to c) and adjacent continuum (panel d) as a function of Doppler velocity, as indicated by labeled points in panels e to g.  The brightness contrast between star and faintest parts shown is $10^5$. Gravity darkening is not very pronounced at this wavelength and as well the equator is ``back-illuminated'' by the disk in this model.
{Middle row:} The resulting spectral profile (panel e) and the interferometric squared visibility (panel f) and phase (panel g) for a baseline along the major axis ($x$-direction), computed for a spectral resolving power of $R=12\,000$.
{Lower row:} Percentage of integrated circumstellar flux as a function of distance from the center along the major axis, i.e., as seen by an interferometric instrument, for $K$-band, \HA, and Br$\gamma$ for an intermediate (panel h) and high disk base density (panel i). The hatched regions indicate the range from the (smaller) pole-on to (larger) edge-on cases.
The dotted limit marks 80\% of the total flux, corresponding to the FWHM for a Gaussian shaped emission profile. The inclination angle for panels a to g is 45\deg, for panels h and i the full range from 0\deg to 90\deg is covered.
From \citet{2013A&ARv..21...69R}. 
%

}

\end{figure*}

For a Be disk of typical density, the midplane temperature  of the disk quickly drops with increasing distance from the stellar surface, as it is optically thick and hence shielded from the stellar radiation. With decreasing opacity, it rises back to temperatures of maybe half to two thirds of that of the photosphere. This happens both in the radial as well as in the vertical direction (subpanels b and c). Close to the star, given sufficient density, this structure of the VDD will have the effect to create a "pseudo-photosphere" that contributes to the SED and becomes more extended with longer wavelengths. 
The re-processing of light in the visible outer layers of that non-spherical pseudo-photosphere is the main reason for the SED brightening due to the disk, and is also the cause of the linear polarization of Be stars. Due to the axial symmetry, the observable polarization is zero for pole-on Be stars, then increases to a maximum at between 70 and 80\deg inclination, and then drops again, as the polarized light is self-absorbed by the now edge-on seen disk. In a given star, the polarization is typically stable in angle and varies in intensity with the base density of the disk. Only during mass ejection in an outburst some small jittering of these parameters has been observed.
How the structure of a steady state VDD translates into photometric, spectroscopic and interferometric observables is illustrated in Fig.~\ref{fig:lineform}. 

An important, somewhat non-intuitive consequence of the VDD is that only a fraction of the initially ejected material leaves the system. 
While the net flow of gas and AM will be outwards as long as the disk is fed, on a local scale material moves both in- and outwards. The gas particles exchange AM through viscosity, propelling a small fraction to higher orbits, but the majority will ultimately fall back to the star. In a steady state disk, this is balanced by the continuous injection of new material and AM from below, but in a disk with episodic feeding this can cause strong deviations from the steady state configuration of the VDD.

\subsection{Disk variability}\label{subsec:diskvar}

For a Be star without a prior disk, i.e., one that is forming a new disk, the orbits of the newly ejected particles are circularized within a few days, leading to a symmetric disk close to the star. If the mass injection continues, the viscosity will then lead to a slowly outwards growing disk where the inner part approaches the steady-state configuration described above earlier. As a consequence, the density distribution in a forming disk is radially steeper than the steady state value of $r^{-3.5}$. If the mass ejection does not continue, this protodisk decays again quickly, and this leads to a phenomenon that has been dubbed "flickering", where weak emission and moderate photometric brigthening build up in a matter of days to weaks, and then disappear over weeks to months. 

If the mass ejection goes on, however, a Be star with a newly forming disk will brighten as the pseudo-photosphere increases in size. Since the pseudo-photosphere in the steady state is more extended for longer wavelengths, this leads to a reddening of the SED of the Be star vs.\ the central object alone as the disk is formed.
Conversely, when the disk is not replenished anymore and starts to decay, this also happens inwards-out, but now is marked by a shallower density law, as the inner material returns to the star more quickly. In such a decaying disk, the net flow of gas and AM is reversed as the star now re-accretes the previously ejected gas. 

Since the timescales for a complete disk build up and decay can be very long, no reliable statistics are yet available, but for an increasing number of objects it is observed that after a period of flickering activity for a few years they form a strong disk again over some years, which then may remain stable for a long time, and finally fully decay over one or two decades. After a time of complete quiescence, this activity cycle may start again. That the disk is built up faster than the subsequent decay is a general property observed in Be star disks, and the understanding gained from the VDD concept has enabled to use such observations to estimate the numerical value of the viscosity parameter $\alpha$. It is found that the value is quite large, in the range from 0.25 to 0.65, and can differ in the various phases, with the higher values seen during build-up.

Once the disk has reached at least moderate size, the non-spherical terms in the stellar gravity field, as well as disturbance from binarity\footnote{Phase locked tidally induced effects are discussed in Sect.~\ref{subsec:curbin}}, together with the viscosity, can lead to the formation of a one-armed, slightly spiral shaped density wave in the disk. Such a wave consists of co-aligned, non-spherical orbits of the gas particles, that can precess around the central star, with a full precession cycle taking from a few years up to well more than a decade. The V/R cycle lengths are independent from any binary period. Depending on the azimuthal orientation of the wave they cause strongly asymmetric emission line profiles, where the typically double-peaked appearance shows one peak much stronger than the other. This behaviour has been named cyclic $V/R$ variability. A scheme of such a density wave, together with spectroscopic example profiles, is shown in Fig.~\ref{fig:VR}. In some stars the UV wind lines show a clear sign of stronger radiative ablation at the position of the density wave. A once established density wave can also subside after some cycles and the disk returns to an axisymmetric state again. Considering that the phenomenon is transient over long timescales, it is hard to give a statistics, but in snapshot studies about \sfrac{1}{3} of Be stars with well developed disks show $V/R$ type asymmetries.

While for the majority of Be stars the disk is always well aligned with the stellar equator, there are stars where this is temporarily not the case. The best known examples include $\gamma$\,Cas and Pleione. For Pleione this is largely understood as nodal precession with a period of about 81 years, and the driving force might be the object's binarity. The amplitude of the precession is different between the inner and outer part of the disk, leading to a periodic "tearing" of  the disk \citep{2024MNRAS.527.7515S}. 
Figure~\ref{fig:shell} shows a sequence of Pleione spectra over almost 20 years, from a nearly face-on to a nearly egde-on outer disk. Since the central star does not change, and the disk base density is largely invariant during these years, Pleione also serves as a good illustration of the impact of the disk inclination on the spectral appearance. For $\gamma$\,Cas this scenario can not be applied, since the changes, observed only in the 1940s, were not regular enough to attribute to precession. The responsible mechanism for $\gamma$\,Cas, and an increasing number of fainter Be stars showing the same phenomenon, remains under discussion \citep{2023A&A...678A..47B}.


\section{Binarity}\label{sec:bin}%

Binarity is invoked neither in the broad observational definition of Be stars, nor in the modern definition based on the inferred physical properties (Sect.~1), and binarity is clearly not the direct cause of the `Be phenomenon' that forms the circumstellar disks in Be stars \citep[not even in highly eccentric systems, see][and below]{2013A&ARv..21...69R}. However, it is established that a large proportion ($>50$\%) of ZAMS B-type stars are found in binary systems whose components will interact during their respective lifetimes, and the role of past binary interactions in Be stars turns out to be crucial as well. In fact, at least a significant fraction of Be stars are in a post-mass-transfer stage of close binary evolution \citep{2021AJ....161..248W}. Importantly, past binary interactions offer an alternative to the evolutionary spin-up of the present-day Be star, as the needed angular momentum could be gained from the orbiting companion instead (Sect. 4.1). 

Some semidetached mass-transferring binaries, whose hydrogen emission line morphology may closely resemble that of Be stars, and which were in the past suggested to represent Be stars at large, may thus in fact be immediate progenitors of Be stars, but with accretion disks. 
The orbital periods of these binaries are typically of the order of days to weeks and may include a certain subset of (non-eclipsing) Algol-types with hot mass gainers such as W~Ser-type binaries and the class of double-periodic variables.

\subsection{Be stars as binary products}\label{subsec:binprod}

The binary evolution hypothesis of the formation of Be stars \citep{1991A&A...241..419P} can be summarized as follows. Suppose a binary star is composed of two  early-type stars in a close orbit with a period of days to weeks. The initially more massive component, that will be the mass donor and future stripped star, evolves off the main sequence, filling its Roche lobe and initiating a steady mass and angular momentum transfer to the less massive component, the mass gainer and future Be star. This phase leads to the expansion of the orbit, now with periods of months, and mass ratio reversal with the mass donor losing most of its hydrogen envelope, revealing its helium-burning core. The rejuvenated mass gainer spins up and forms a self-ejected Keplerian disk to become a Be star. Immediately following the mass transfer, the mass donor for a brief time ($\sim 0.1$ to $1$\,Myr) appears as a `bloated' stripped star. In this phase, the stripped star retains the spectral appearance of a main sequence early-type star, both in terms of effective temperature and luminosity, but its mass already has become only a small fraction of the initial one. As the stripped star shrinks and settles into equilibrium, its temperature increases while the luminosity decreases, after which the newly formed Be star dominates the observed spectrum, potentially hiding the signature of the stripped companion entirely. The stripped companion continues the nuclear burning of helium, which may be followed by a late stage of helium shell-burning associated with increased effective temperatures. If massive enough, the stripped star (referred to as a subdwarf OB-type star\footnote{The general population of subdwarf O- and B-type stars resides between the upper main sequence and the white dwarf sequence, on the so-called extreme horizontal branch in the Hertzsprung-Russell diagram, and is probably dominated by late evolutionary stages of low-mass stars.} or a helium star) can subsequently explode in a supernova, forming a neutron star or possibly a black hole, while a stripped star of low mass could evolve to become a white dwarf, if the associated timescale is shorter than that of the Be star meanwhile evolving into a supergiant. A third, as of now only theoretically suggested possibility is that the helium star companion might enter a second mass transfer as it evolves, called case BB, which might lead to a Be plus white dwarf system much more quickly than without such a secondary mass transfer.

Over the recent decades, studies employing binary population synthesis methods to investigate the origin of Be stars have consistently concluded that the binary formation channel is of great importance and possibly even dominates the evolution of slowly rotating B-type stars into rapidly rotating Be stars that can form disks \citep{2014ApJ...796...37S}. The predictions for the present-day Galactic population of Be stars formed in binary evolution are such that Be stars are commonly expected in binaries with different types of stripped companions, depending on the initial masses and orbital period of the progenitor binary. White dwarf and subdwarf OB-type (sdOB) companions should be particularly common, while Be stars with neutron star and black hole companions should be rare. In addition to Be stars that remain bound binaries after the mass-transfer phase, Be stars may also originate from stellar mergers, although the physics of this process are still poorly understood. Finally, some Be stars whose stripped companions explode in supernovae may become unbound, so that the now single Be star may become observable as a `runaway' star with a peculiarly high space velocity \citep{2001ApJ...555..364B}. 


\subsection{Be stars as current binaries}\label{subsec:curbin}

Observational efforts have revealed a picture consistent with many Be stars having been spun up in past binary interactions, although the role and prominence of evolutionary spin-up is still highly debated. Nevertheless, the fact is that among bright and well-observed Be stars, the fraction of confirmed multiple systems is about $20-30$\%. In terms of orbital period distribution, Be stars should generally not exist in binaries with periods shorter than a few weeks, as the companions would prevent the build-up of the Be disks, and this is indeed compatible with the observations\footnote{The shortest well-constrained orbital period is $28.2$ days for 59~Cyg. There are several BeXRBs
with periods as short as two weeks, but orbital misalignment may play a role in allowing a disk to form in systems with such short periods.}. There are currently no confirmed Be stars with main sequence companions with orbital periods of less than a few years\footnote{The shortest well-constrained orbital period is $7.04$ years for Achernar.}; the companions that can be identified in this shorter period range are in all cases stripped, evolved stars (see below). As main-sequence binaries with periods of the order of months to a few years are common among B-type stars, the lack of such binaries in the Be population alone gives a strong support to the binary origin hypothesis.  

The entire class of Be X-ray binaries (a subset of high-mass X-ray binaries numbering $\sim70$ objects in the Galaxy\footnote{According to the online catalog maintained at \url{http://xray.sai.msu.ru/~raguzova/BeXcat/}.}) clearly originates from past binary mass transfer, after which the stripped companions remained massive enough and eventually exploded in supernovae \citep{2011Ap&SS.332....1R}. The compact object can then accrete material from the Be disk, giving rise to conspicuous non-thermal X-ray emission, which enables the detection of these rare objects even at large distances. Thus far, only neutron stars were confirmed as the compact companions, while no black hole companions are currently known. In addition, there are a few sources of supersoft X-ray emissions that were proposed to be Be stars paired with white dwarf.

A controversial subclass of X-ray sources associated with Be stars - currently containing about 25 confirmed cases - is exemplified by the unusually hard but thermal X-rays detected from the early binary Be star $\gamma$~Cas \citep[the subclass is sometimes referred to as `$\gamma$~Cas analogs', although the similarity between the objects pertains solely to the type of X-rays and early-B spectral subtypes,][]{2016AdSpR..58..782S}. The X-ray luminosity for these stars is lower compared to Be X-ray binaries ($\log{L_{\rm X}} \sim 31.6$ to $33.2$, $\log{L_{\rm X}/L_{\rm BOL}} \sim -6.2$ to $-4.0$), while the X-ray hardness is higher than for the X-rays from stellar winds (hardness ratio, i.e., the ratio of hard to soft X-rays, corresponding to energy bands of 2--10\,keV and 0.5--2\,keV, respectively, is $>1.6$, while plasma temperature $kT>5$\,keV, and $L_{\rm X, hard} > 31$). A third property is variability of the X-ray flux on all timescales. Accretion of Be disk material onto white dwarf companions (which could be via a disk if the white dwarf is non-magnetic or via accretion columns if it is magnetic) has been invoked to explain the X-ray formation, although a debate is still ongoing in regards to another suggested scenario, in which the X-rays are speculated to originate from (small scale) magnetic field reconnections between the stellar surface and the inner parts of the Be disks and have nothing to do with binarity. However, in addition to no magnetic field of either scale having yet been detected in Be stars, there is a growing amount of evidence indicating that Be stars with $\gamma$~Cas-like X-rays are generally binaries with invisible companions that are neither subdwarf OB-type stars nor neutron stars. In fact, signatures of binarity - specifically radial velocity variations of the Be spectral lines consistent with orbits of the order of months - have already been found for about a half of these stars \citep{2022MNRAS.510.2286N}. The best studied binary in this group is $\gamma$~Cas itself, for which a white dwarf companion seems to be the only possible option by process of elimination, as other types of companions for the spectroscopically well-constrained mass of $\sim1$\,M$_\odot$ were recently ruled out with interferometric observations.  

Moving on from compact companions and associated X-rays, we enter the class of Be stars with subdwarf OB-type (sdOB) companions, which are difficult to detect due to their faintness. About two dozen confirmed Be + sdOB systems with orbital periods of the order of months are known among bright and nearby Be stars, with the number steadily growing in recent years \citep{2021AJ....161..248W}. The sdOB companions are typically hotter than the Be stars, and this enables their detection in the far UV (FUV) spectral domain, where they have been found to contribute up to $\sim15$\% of the observed flux. Conversely, these objects do not display any X-ray properties that would set them apart. The first sdOB companion was confirmed for the Be star $\varphi$~Per in FUV spectra from the HST, and the astrometric orbit was subsequently mapped with interferometry in the near-IR, where the companion in this case contributes $\sim1.5$\% of the total flux \citep{2015A&A...577A..51M}. These observations enabled the first model-independent characterization of a Be star binary system including a full orbital solution with dynamical masses and parallax, and $\varphi$~Per thus became the first unbiased anchor point for binary evolution models leading to the formation of Be stars. Since then, more sdOB companions were found in archival and new FUV spectra, and more orbits were mapped by interferometry, making this subclass of Be binaries the only one with a significant set of fundamental parameters available.

The fraction of Be stars that are interacting binary products appears to be dependent on the spectral subtype, with the early-type Be stars being the most commonly identified as post-mass-transfer binaries, and with the fraction rapidly dropping towards the later subtypes. This may be due to the evolutionary spin-up formation channel of Be stars becoming more prominent towards later subtypes. Nevertheless, the case of the B6e star $\kappa$~Dra shows that at least some mid- to late-subtypes do have stripped companions, albeit of the cooler sdB type.

The short-lived evolutionary stage immediately following the binary mass transfer and leading to the formation of the Be + sdOB binaries - which was referred to above as Be stars with `bloated' stripped companions - was recently observationally confirmed for the first time \citep{2020A&A...641A..43B}. These systems are extremely important for stellar astrophysics, as they represent one of the most physically interesting links in Be binary evolution, which was missing until now. Statistics of a representative sample of Be stars following the discovery of several of these objects suggest that there might be around five sdO companions for each `bloated' star, all of them found around early type Be stars.

There are other binary systems containing Be stars which are not as straightforward to interpret on the basis of binary evolution. Among these, there are stars that are orbited by likely main sequence companions on highly eccentric orbits with orbital periods of the order of years to decades, examples of which include the well-known Be stars Achernar and $\delta$~Sco. The origin of the rapid rotation in these systems is uncertain and at least in the case of Achernar, strong arguments have been presented in favor of the evolutionary spin-up scenario, as the architecture of the system does not leave room for a stripped companion close to the Be star \citep{2022A&A...667A.111K}. In $\delta$~Sco, the Be nature began to manifest itself only during one of the recent periastron passages, leading to suggestions that the gravitational pull of the passing companion could have provided the needed additional force leading to disk ejection. However, this was later ruled out as the disk actually started to form already before the periastron passage, while the pull from the companion was found to be insufficient to significantly influence the gravitational potential near the Be star.

Be stars are also found in higher-order multiple systems. Among these, there are several known objects in which the Be star is the outer component in a hierarchical triple system, with the inner binary composed of two stars on a short-period and nearly-circular orbit, and with the wide orbit typically being eccentric \citep{2021ApJ...916...24K}. The evolutionary origin of the Be stars in these systems is unclear, as no additional close stripped companions to the Be stars have been revealed thus far. Theoretical scenarios include the remnantless explosion of the former donor, and the origin of the Be star in a stellar merger, caused by Lidov-Kozai oscillations excited by the companion on a wide eccentric orbit (in this case also a binary star). However, no evidence has been presented yet to support this scenario, and evolutionary spin-up also remains a viable option for the Be stars in these possibly primordial systems. 

An example of a peculiar multiple system containing a Be star is 59~Cyg; it is at least a quadruple system composed of an inner (slightly eccentric) Be + sdO binary (period of 28\,d), with a third component on a highly eccentric and probably misaligned orbit (preliminary period of 906\,d), and with a fourth component on a very wide orbit \citep[preliminary period of 161.5\,yr][]{2024ApJ...962...70K}. The third component could be responsible for the highly untypical non-zero eccentricity of the inner Be + sdO pair\footnote{Only one other Be + sdO binary -- 60~Cyg -- is currently known to be slightly eccentric.}.

\subsection{Influence of companions on the disk}\label{subsec:influence}

There might also be Be stars whose close companions are on orbits misaligned with the Be disk plane. This is probably the case for the Be star Pleione, in which observations and modeling suggest that the outer part of its disk gets misaligned with and ultimately detaches from the inner part of the disk. After the tearing, the orbital plane of the outer disk proceeds to precess around the stellar rotation axis before dissipating, as it is no longer supplied with AM and matter from the inner disk \citep{2022ApJ...928..145M}. Pleione's companion of unknown type is on a 218-day orbit which was also reported to be highly eccentric. Unfortunately, no further details have been presented about the binary orbit, as the companion remains undetected. Other Be stars have shown signs of tilting and possibly precessing disks, but details about the suspected misalignment of their binary orbits are not available as of yet. 

In the majority of binary Be stars, the orbital and disk planes are aligned. Among these, the tidal influence of a companion orbiting a Be star with a developed disk gives rise to density waves propagating through the disk with the orbital period. A two-armed spiral density wave is expected to form, although the arm directly facing the companion can be much stronger than the other one. This can be observed as a orbital phase-locked small-scale variation in the $V/R$ ratio of strong emission lines. Another possibility is that the companion influences the vertical structure of the disk, so that a `puffed up' region may pass in front of the star, resulting in orbital phase-locked transitions between a normal Be and a Be-shell type spectral appearance.

Hot companions such as stripped sdOB stars also have a radiative influence on the Be disk, and the extra ionization in the outer part of the Be disk facing the companion gives rise to observable features, again phased with the orbital period. The typical observed feature is in this case an extra emission peak in singly ionized helium lines that moves with the orbital period, but with larger velocities than the companion itself. This is because this extra emission originates from the outer parts of the Be disk, where the Keplerian velocities are higher than for the companion. This feature can be quite prominent, and many of the currently known Be + sdOB binaries were in fact first suspected from its detection in optical spectra.


\section{Conclusion}\label{sec:concl}%

While line emission arises easily when gas is present in the circumstellar environment of any B star, at least for main sequence B-type stars, most objects showing such emission are classical Be stars. These objects are common, up to 20\% of all main sequence B stars are Be stars, with an increasing fraction towards the end of the MS life time. For a long time, observations indicated there to be more early type Be stars than late type ones, but better data suggest this to be an observational selection effect because the late subtypes have weaker and more tenuous disks, but disks nonetheless. These disks are in Keplerian rotation around the host star, and formed from material ejected from the stellar photosphere. The behaviour of the disk is mostly governed by viscosity, and in binaries additionally by tidal effects. At higher latitudes, i.e., the vertically outer disk regions, for tenuous disks, and for very early B subtypes radiative ablation becomes significant as well.

In stellar evolutionary terms, the outflowing disk can be considered a "spillover" effect, where, on top of the rotation, the mass and AM injection from the stellar surface into the disk is driven by additional mechanisms, like the ubiquitous nonradial pulsation, and possibly others. Such a disk is a means to shed, and hence regulate, the angular momentum of the stellar surface to stay below the critical value. For early B subtypes, Be stars acquire this high AM predominantly through binary interaction, but for later subtypes the internal stellar evolution may also play a role in driving the stellar surface towards its critical rotation value $\omeorb \rightarrow 1$. As soon as the conditions, i.e., rotation plus additional mechanism, allow the formation of a disk and thus to shed AM, it will happen. The internal AM evolution alone may then be sufficient to keep all Be stars at least temporarily active until the MS turn-off, so that the Be fraction increases towards the end of the core hydrogen burning phase. A VDD is highly efficient in removing AM from the system, meaning it does so with only little overall mass loss: Most of the initially ejected material will transfer AM outwards, but the matter itself will eventually fall back to the star.

As the short term stellar variability, like NRP, is typically stronger in earlier subtypes, disks are more easily formed in those, i.e. the Be phenomenon sets in at lower critical fractions \omeorb. The disks tend to be more dense and variable in early subtypes, and more steady-state and tenuous in late ones. This might be a consequence of the longer MS lifetimes and slower internal AM evolution, and less intense radiative disk ablation in late B-subtypes.

Many Be stars, in particular among the early subtypes, are known to have acquired their rapid rotation in binary interaction, as is evident from their evolved companions. Theoretically, also single stars can become Be stars if they begin their MS life as sufficiently rapid rotators. While no individual object could yet be conclusively proven to originate from this internal evolutionary formation channel, the statistics of evolved companions suggests single star evolution to be an increasingly relevant formation mechanism towards later type Be stars.
For fully formed Be stars binarity remains relevant, as the companion may exert an influence on the disk, both through gravitational and radiative interaction. In Be stars with compact companions, accretion of outwardly migrating disk material onto the companion may be visible in the high energy wavelength range, giving rise to the class of Be X-ray binaries, BeXRB.

The general occurence of Be stars does, in principle, not depend on metallicity (which may affect the presence of NRP), since mass ejection mechanisms will eventually become available to any rapidly rotating star to form a Keplerian disk, whether they be NRP, tidal stripping, rotation and turbulence alone, or something else entirely.
Metallicity may, however, affect the general incidence of Be stars through the differences in general AM evolution and mass loss, as well as the threshold value of \omeorb at which Be stars will begin to form. The potential of extragalactic Be stars, several hundreds of which are already well known for the Magellanic Clouds, is yet to be fully exploited, however, since long-term observations of a large number of objects are vital to fully understand statistics and processes of Be stars as a group. The next generation of photometric and spectroscopic surveys will enable progress here.

With all the above in mind, the biggest remaining questions on Be stars are: 
\begin{itemize}
    \item The details of the, possibly, various formation channels and the internal AM evolution of the central star.
    \item The physics of the disk ejection mechanism and related to this, the pulsation physics of Be stars. Asteroseismology promises great potential for this and the first point once rapid rotation and nonlinear effects have been included.
    \item Whether Be stars in their post-MS evolution produce different outcomes than non Be stars.
    \item The nature and fate of the companions in the different binary scenarios. 
    \item As a testbed for the investigation astrophysical disks in general, since Be disks are among the brightest and closest examples. The timescales for the disk build up and decay cycles can be very long and Be stars may temporarily pose as B stars, meaning Be stars have a "duty cycle" of less than 100\%. Therefore, only now after several decades of nearly continuous observation of a large number of Be stars in professional and amateur projects, a statistical understanding can be formed.
    \item How extragalactic Be stars, at different metallicity, behave different from Galactic ones.
\end{itemize}

 For further reading, more detailed works and reviews can be found on 
 classical Be stars by \citet{1988PASP..100..770S,2003PASP..115.1153P,2013A&ARv..21...69R}, 
 the rotation of active B stars by \citep{2023Galax..11...54Z},
 Be X-ray binaries by \citet{2011Ap&SS.332....1R},
 specifically $\gamma$\,Cas analogs by \citet{2016AdSpR..58..782S},
 B[e] stars by \citet{2007ApJ...667..497M}, and
 Herbig Ae/Be stars by \citet{1998ARA&A..36..233W}. 
 Reviews on wider, but related issues can be found about
 the rotation of massive stars by \citet{2013ApJ...764..166D}, 
 early type stars with magnetospheres by \citet{2013MNRAS.429..398P}, 
 the mass-loss of hot massive stars in general by \citet{2008A&ARv..16..209P}, and 
 the hydrodynamical aspects of those winds in particular by \citet{2023Galax..11...68C}.
Meeting and conference proceedings that were entirely about or had at sessions on Be stars, summarizing the state of research, can be found in \citet{2016ASPC..506.....S}, \citet{2017ASPC..508.....M},   \citet{2017IAUS..329.....E}, and \citet{2020svos.conf.....N}. 


\begin{ack}[Acknowledgments]
~We thank Dietrich Baade, 
Carol Jones, Douglas Gies, and Yael Naz\'e for comments on the manuscript. 
The study of Be stars benefits greatly from the policy of openness that most astrophysical observatories and institutions have adopted for their data as well as their publications. 
This includes the amateur observers contributing to the BeSS database, operated at LESIA, Observatoire de Meudon, France: {\tt http://basebe.obspm.fr}, from which some of the spectra shown in the figures were drawn.
\end{ack}

\seealso{Asteroseismology
-- Spectral classification of stars
-- Evolution of binary stars
-- Stellar rotation
-- Stellar winds
-- Blue straggler stars
-- Hot subdwarf stars
-- High-mass X-ray Binaries
}

\bibliographystyle{Harvard}
\begin{thebibliography*}{48}
\providecommand{\bibtype}[1]{}
\providecommand{\natexlab}[1]{#1}
{\catcode`\|=0\catcode`\#=12\catcode`\@=11\catcode`\\=12
|immediate|write|@auxout{\expandafter\ifx\csname
  natexlab\endcsname\relax\gdef\natexlab#1{#1}\fi}}
\renewcommand{\url}[1]{{\tt #1}}
\providecommand{\urlprefix}{URL }
\expandafter\ifx\csname urlstyle\endcsname\relax
  \providecommand{\doi}[1]{doi:\discretionary{}{}{}#1}\else
  \providecommand{\doi}{doi:\discretionary{}{}{}\begingroup
  \urlstyle{rm}\Url}\fi
\providecommand{\bibinfo}[2]{#2}
\providecommand{\eprint}[2][]{\url{#2}}

\bibtype{Inproceedings}%
\bibitem[{Baade} and {Balona}(1994)]{1994IAUS..162..311B}
\bibinfo{author}{{Baade} D} and \bibinfo{author}{{Balona} LA}
  (\bibinfo{year}{1994}), \bibinfo{title}{{Periodic Variability of Be Stars:
  Nonradial Pulsation or Rotational Modulation?}}, \bibinfo{editor}{{Balona}
  LA} et~al, (Eds.), \bibinfo{booktitle}{Pulsation; Rotation; and Mass Loss in
  Early-Type Stars}, \bibinfo{series}{IAU Symposium}, \bibinfo{volume}{162},
  pp. \bibinfo{pages}{311}.

\bibtype{Article}%
\bibitem[{Baade} et al.(2023)]{2023A&A...678A..47B}
\bibinfo{author}{{Baade} D} et~al (\bibinfo{year}{2023}).
\bibinfo{title}{{The historical active episodes of the disks around
  {\ensuremath{\gamma}} Cassiopeiae (B0.5 IVe) and 59 Cygni (B1 IVe)
  revisited}}.
\bibinfo{journal}{{\em \aap}} \bibinfo{volume}{678}, \bibinfo{eid}{A47}.
  \bibinfo{doi}{\doi{10.1051/0004-6361/202244149}}.

\bibtype{Article}%
\bibitem[{Berger} and {Gies}(2001)]{2001ApJ...555..364B}
\bibinfo{author}{{Berger} DH} and \bibinfo{author}{{Gies} DR}
  (\bibinfo{year}{2001}).
\bibinfo{title}{{A Search for High-Velocity Be Stars}}.
\bibinfo{journal}{{\em \apj}} \bibinfo{volume}{555}: \bibinfo{pages}{364--367}.
  \bibinfo{doi}{\doi{10.1086/321461}}.

\bibtype{Article}%
\bibitem[{Bjorkman} and {Cassinelli}(1993)]{1993ApJ...409..429B}
\bibinfo{author}{{Bjorkman} JE} and \bibinfo{author}{{Cassinelli} JP}
  (\bibinfo{year}{1993}).
\bibinfo{title}{{Equatorial Disk Formation around Rotating Stars Due to Ram
  Pressure Confinement by the Stellar Wind}}.
\bibinfo{journal}{{\em \apj}} \bibinfo{volume}{409}: \bibinfo{pages}{429}.
  \bibinfo{doi}{\doi{10.1086/172676}}.

\bibtype{Article}%
\bibitem[{Bodensteiner} et al.(2020)]{2020A&A...641A..43B}
\bibinfo{author}{{Bodensteiner} J} et~al (\bibinfo{year}{2020}).
\bibinfo{title}{{Is HR 6819 a triple system containing a black hole?. An
  alternative explanation}}.
\bibinfo{journal}{{\em \aap}} \bibinfo{volume}{641}, \bibinfo{eid}{A43}.
  \bibinfo{doi}{\doi{10.1051/0004-6361/202038682}}.

\bibtype{Inproceedings}%
\bibitem[{Carciofi} et al.(2017)]{2017IAUS..329..390C}
\bibinfo{author}{{Carciofi} AC} et~al (\bibinfo{year}{2017}),
  \bibinfo{title}{{HDUST3 - A chemically realistic, 3-D, NLTE radiative
  transfer code}}, \bibinfo{editor}{{Eldridge} JJ} et~al, (Eds.),
  \bibinfo{booktitle}{The Lives and Death-Throes of Massive Stars},
  \bibinfo{series}{IAU Symposium}, \bibinfo{volume}{329},
  \bibinfo{pages}{390--390}.

\bibtype{Article}%
\bibitem[{Cassinelli} et al.(2002)]{2002ApJ...578..951C}
\bibinfo{author}{{Cassinelli} JP} et~al (\bibinfo{year}{2002}).
\bibinfo{title}{{A Magnetically Torqued Disk Model for Be Stars}}.
\bibinfo{journal}{{\em \apj}} \bibinfo{volume}{578}: \bibinfo{pages}{951--966}.
  \bibinfo{doi}{\doi{10.1086/342654}}.

\bibtype{Article}%
\bibitem[{Cur{\'e}} and {Araya}(2023)]{2023Galax..11...68C}
\bibinfo{author}{{Cur{\'e}} M} and \bibinfo{author}{{Araya} I}
  (\bibinfo{year}{2023}).
\bibinfo{title}{{Radiation-Driven Wind Hydrodynamics of Massive Stars: A
  Review}}.
\bibinfo{journal}{{\em Galaxies}} \bibinfo{volume}{11}, \bibinfo{eid}{68}.
  \bibinfo{doi}{\doi{10.3390/galaxies11030068}}.

\bibtype{Article}%
\bibitem[{de Mink} et al.(2013)]{2013ApJ...764..166D}
\bibinfo{author}{{de Mink} SE} et~al (\bibinfo{year}{2013}).
\bibinfo{title}{{The Rotation Rates of Massive Stars: The Role of Binary
  Interaction through Tides, Mass Transfer, and Mergers}}.
\bibinfo{journal}{{\em \apj}} \bibinfo{volume}{764}, \bibinfo{eid}{166}.
  \bibinfo{doi}{\doi{10.1088/0004-637X/764/2/166}}.

\bibtype{Proceedings}%
\bibitem[Eldridge et al.(2017)]{2017IAUS..329.....E}
\bibinfo{editor}{Eldridge JJ} et~al, (Eds.) (\bibinfo{year}{2017}).
\bibinfo{title}{{The Lives and Death-Throes of Massive Stars}},
  \bibinfo{comment}{vol.} \bibinfo{volume}{329}, \bibinfo{series}{IAU
  Symposium}.
\bibinfo{doi}{\doi{10.1017/S1743921317003623}}.
\bibtype{Article}%
\bibitem[{Grady} et al.(1996)]{1996ApJ...471L..49G}
\bibinfo{author}{{Grady} CA} et~al (\bibinfo{year}{1996}).
\bibinfo{title}{{The beta Pictoris Phenomenon in A-Shell Stars: Detection of
  Accreting Gas}}.
\bibinfo{journal}{{\em \apjl}} \bibinfo{volume}{471}: \bibinfo{pages}{L49}.
  \bibinfo{doi}{\doi{10.1086/310332}}.

\bibtype{Article}%
\bibitem[{Kee} et al.(2016)]{2016MNRAS.458.2323K}
\bibinfo{author}{{Kee} ND} et~al (\bibinfo{year}{2016}).
\bibinfo{title}{{Line-driven ablation of circumstellar discs - I. Optically
  thin decretion discs of classical Oe/Be stars}}.
\bibinfo{journal}{{\em \mnras}} \bibinfo{volume}{458}:
  \bibinfo{pages}{2323--2335}. \bibinfo{doi}{\doi{10.1093/mnras/stw471}}.

\bibtype{Article}%
\bibitem[{Kervella} et al.(2022)]{2022A&A...667A.111K}
\bibinfo{author}{{Kervella} P} et~al (\bibinfo{year}{2022}).
\bibinfo{title}{{The binary system of the spinning-top Be star Achernar}}.
\bibinfo{journal}{{\em \aap}} \bibinfo{volume}{667}, \bibinfo{eid}{A111}.
  \bibinfo{doi}{\doi{10.1051/0004-6361/202244009}}.

\bibtype{Article}%
\bibitem[{Klement} et al.(2019)]{2019ApJ...885..147K}
\bibinfo{author}{{Klement} R} et~al (\bibinfo{year}{2019}).
\bibinfo{title}{{Prevalence of SED Turndown among Classical Be Stars: Are All
  Be Stars Close Binaries?}}
\bibinfo{journal}{{\em \apj}} \bibinfo{volume}{885}, \bibinfo{eid}{147}.
  \bibinfo{doi}{\doi{10.3847/1538-4357/ab48e7}}.

\bibtype{Article}%
\bibitem[{Klement} et al.(2021)]{2021ApJ...916...24K}
\bibinfo{author}{{Klement} R} et~al (\bibinfo{year}{2021}).
\bibinfo{title}{{{\ensuremath{\nu}} Gem: A Hierarchical Triple System with an
  Outer Be Star}}.
\bibinfo{journal}{{\em \apj}} \bibinfo{volume}{916}, \bibinfo{eid}{24}.
  \bibinfo{doi}{\doi{10.3847/1538-4357/ac062c}}.

\bibtype{Article}%
\bibitem[{Klement} et al.(2024)]{2024ApJ...962...70K}
\bibinfo{author}{{Klement} R} et~al (\bibinfo{year}{2024}).
\bibinfo{title}{{The CHARA Array Interferometric Program on the Multiplicity of
  Classical Be Stars: New Detections and Orbits of Stripped Subdwarf
  Companions}}.
\bibinfo{journal}{{\em \apj}} \bibinfo{volume}{962}, \bibinfo{eid}{70}.
  \bibinfo{doi}{\doi{10.3847/1538-4357/ad13ec}}.

\bibtype{Article}%
\bibitem[{K\v{r}\'i\v{z}} and {Harmanec}(1975)]{1975BAICz..26...65K}
\bibinfo{author}{{K\v{r}\'i\v{z}} S} and \bibinfo{author}{{Harmanec} P}
  (\bibinfo{year}{1975}).
\bibinfo{title}{{A Hypothesis of the Binary Origin of Be Stars}}.
\bibinfo{journal}{{\em Bulletin of the Astronomical Institutes of
  Czechoslovakia}} \bibinfo{volume}{26}: \bibinfo{pages}{65}.

\bibtype{Article}%
\bibitem[{Labadie-Bartz} et al.(2022)]{2022AJ....163..226L}
\bibinfo{author}{{Labadie-Bartz} J} et~al (\bibinfo{year}{2022}).
\bibinfo{title}{{Classifying Be Star Variability With TESS. I. The Southern
  Ecliptic}}.
\bibinfo{journal}{{\em \aj}} \bibinfo{volume}{163}, \bibinfo{eid}{226}.
  \bibinfo{doi}{\doi{10.3847/1538-3881/ac5abd}}.

\bibtype{Article}%
\bibitem[{Marr} et al.(2022)]{2022ApJ...928..145M}
\bibinfo{author}{{Marr} KC} et~al (\bibinfo{year}{2022}).
\bibinfo{title}{{The Role of Disk Tearing and Precession in the Observed
  Variability of Pleione}}.
\bibinfo{journal}{{\em \apj}} \bibinfo{volume}{928}, \bibinfo{eid}{145}.
  \bibinfo{doi}{\doi{10.3847/1538-4357/ac551b}}.

\bibtype{Article}%
\bibitem[{Meilland} et al.(2012)]{2012A&A...538A.110M}
\bibinfo{author}{{Meilland} A} et~al (\bibinfo{year}{2012}).
\bibinfo{title}{{First spectro-interferometric survey of Be stars. I.
  Observations and constraints on the disk geometry and kinematics}}.
\bibinfo{journal}{{\em \aap}} \bibinfo{volume}{538}, \bibinfo{eid}{A110}.
  \bibinfo{doi}{\doi{10.1051/0004-6361/201117955}}.

\bibtype{Article}%
\bibitem[{Miroshnichenko}(2007)]{2007ApJ...667..497M}
\bibinfo{author}{{Miroshnichenko} AS} (\bibinfo{year}{2007}).
\bibinfo{title}{{Toward Understanding the B[e] Phenomenon. I. Definition of the
  Galactic FS CMa Stars}}.
\bibinfo{journal}{{\em \apj}} \bibinfo{volume}{667}: \bibinfo{pages}{497--504}.
  \bibinfo{doi}{\doi{10.1086/520798}}.

\bibtype{Proceedings}%
\bibitem[Miroshnichenko et al.(2017)]{2017ASPC..508.....M}
\bibinfo{editor}{Miroshnichenko A} et~al, (Eds.) (\bibinfo{year}{2017}).
\bibinfo{title}{{The B[e] Phenomenom: Forty Years of Studies}},
  \bibinfo{comment}{vol.} \bibinfo{volume}{508}, \bibinfo{series}{Astronomical
  Society of the Pacific Conference Series}.

\bibtype{Article}%
\bibitem[{Mourard} et al.(2015)]{2015A&A...577A..51M}
\bibinfo{author}{{Mourard} D} et~al (\bibinfo{year}{2015}).
\bibinfo{title}{{Spectral and spatial imaging of the Be+sdO binary
  <ASTROBJ>{\ensuremath{\phi}} Persei</ASTROBJ>}}.
\bibinfo{journal}{{\em \aap}} \bibinfo{volume}{577}, \bibinfo{eid}{A51}.
  \bibinfo{doi}{\doi{10.1051/0004-6361/201425141}}.

\bibtype{Article}%
\bibitem[{Naz{\'e}} et al.(2022)]{2022MNRAS.510.2286N}
\bibinfo{author}{{Naz{\'e}} Y} et~al (\bibinfo{year}{2022}).
\bibinfo{title}{{Velocity monitoring of {\ensuremath{\gamma}} Cas stars reveals
  their binarity status}}.
\bibinfo{journal}{{\em \mnras}} \bibinfo{volume}{510}:
  \bibinfo{pages}{2286--2304}. \bibinfo{doi}{\doi{10.1093/mnras/stab3378}}.

\bibtype{Article}%
\bibitem[{Negueruela} et al.(2004)]{2004AN....325..749N}
\bibinfo{author}{{Negueruela} I} et~al (\bibinfo{year}{2004}).
\bibinfo{title}{{On the class of Oe stars}}.
\bibinfo{journal}{{\em Astronomische Nachrichten}} \bibinfo{volume}{325}:
  \bibinfo{pages}{749--760}. \bibinfo{doi}{\doi{10.1002/asna.200310258}}.

\bibtype{Proceedings}%
\bibitem[Neiner et al.(2020)]{2020svos.conf.....N}
\bibinfo{editor}{Neiner C} et~al, (Eds.) (\bibinfo{year}{2020}).
\bibinfo{title}{{Stars and their variability, observed from space}}.

\bibtype{Article}%
\bibitem[{Okazaki}(2001)]{2001PASJ...53..119O}
\bibinfo{author}{{Okazaki} AT} (\bibinfo{year}{2001}).
\bibinfo{title}{{Viscous Transonic Decretion in Disks of Be Stars}}.
\bibinfo{journal}{{\em \pasj}} \bibinfo{volume}{53}: \bibinfo{pages}{119--125}.
  \bibinfo{doi}{\doi{10.1093/pasj/53.1.119}}.

\bibtype{Article}%
\bibitem[{Owocki} et al.(1996)]{1996ApJ...472L.115O}
\bibinfo{author}{{Owocki} SP} et~al (\bibinfo{year}{1996}).
\bibinfo{title}{{Inhibition of Wind Compressed Disk Formation by Nonradial
  Line-Forces in Rotating Hot-Star Winds}}.
\bibinfo{journal}{{\em \apjl}} \bibinfo{volume}{472}: \bibinfo{pages}{L115}.
  \bibinfo{doi}{\doi{10.1086/310372}}.

\bibtype{Article}%
\bibitem[{Petit} et al.(2013)]{2013MNRAS.429..398P}
\bibinfo{author}{{Petit} V} et~al (\bibinfo{year}{2013}).
\bibinfo{title}{{A magnetic confinement versus rotation classification of
  massive-star magnetospheres}}.
\bibinfo{journal}{{\em \mnras}} \bibinfo{volume}{429}:
  \bibinfo{pages}{398--422}. \bibinfo{doi}{\doi{10.1093/mnras/sts344}}.

\bibtype{Article}%
\bibitem[{Pols} et al.(1991)]{1991A&A...241..419P}
\bibinfo{author}{{Pols} OR} et~al (\bibinfo{year}{1991}).
\bibinfo{title}{{The formation of Be stars through close binary evolution.}}
\bibinfo{journal}{{\em \aap}} \bibinfo{volume}{241}: \bibinfo{pages}{419}.

\bibtype{Article}%
\bibitem[{Porter} and {Rivinius}(2003)]{2003PASP..115.1153P}
\bibinfo{author}{{Porter} JM} and \bibinfo{author}{{Rivinius} T}
  (\bibinfo{year}{2003}).
\bibinfo{title}{{Classical Be Stars}}.
\bibinfo{journal}{{\em \pasp}} \bibinfo{volume}{115}:
  \bibinfo{pages}{1153--1170}. \bibinfo{doi}{\doi{10.1086/378307}}.

\bibtype{Article}%
\bibitem[{Puls} et al.(2008)]{2008A&ARv..16..209P}
\bibinfo{author}{{Puls} J} et~al (\bibinfo{year}{2008}).
\bibinfo{title}{{Mass loss from hot massive stars}}.
\bibinfo{journal}{{\em \aapr}} \bibinfo{volume}{16}: \bibinfo{pages}{209--325}.
  \bibinfo{doi}{\doi{10.1007/s00159-008-0015-8}}.

\bibtype{Article}%
\bibitem[{Quirrenbach} et al.(1997)]{1997ApJ...479..477Q}
\bibinfo{author}{{Quirrenbach} A} et~al (\bibinfo{year}{1997}).
\bibinfo{title}{{Constraints on the Geometry of Circumstellar Envelopes:
  Optical Interferometric and Spectropolarimetric Observations of Seven Be
  Stars}}.
\bibinfo{journal}{{\em \apj}} \bibinfo{volume}{479}: \bibinfo{pages}{477--496}.
  \bibinfo{doi}{\doi{10.1086/303854}}.

\bibtype{Article}%
\bibitem[{Reig}(2011)]{2011Ap&SS.332....1R}
\bibinfo{author}{{Reig} P} (\bibinfo{year}{2011}).
\bibinfo{title}{{Be/X-ray binaries}}.
\bibinfo{journal}{{\em \apss}} \bibinfo{volume}{332}: \bibinfo{pages}{1--29}.
  \bibinfo{doi}{\doi{10.1007/s10509-010-0575-8}}.

\bibtype{Article}%
\bibitem[{Rivinius} et al.(2013)]{2013A&ARv..21...69R}
\bibinfo{author}{{Rivinius} T} et~al (\bibinfo{year}{2013}).
\bibinfo{title}{{Classical Be stars. Rapidly rotating B stars with viscous
  Keplerian decretion disks}}.
\bibinfo{journal}{{\em \aapr}} \bibinfo{volume}{21}, \bibinfo{eid}{69}.
  \bibinfo{doi}{\doi{10.1007/s00159-013-0069-0}}.

\bibtype{Article}%
\bibitem[{Secchi}(1866)]{1866AN.....68...63S}
\bibinfo{author}{{Secchi} A} (\bibinfo{year}{1866}).
\bibinfo{title}{{Schreiben des Herrn Prof. Secchi, Directors der Sternwarte des
  Collegio Romano, an den Herausgeber}}.
\bibinfo{journal}{{\em Astronomische Nachrichten}} \bibinfo{volume}{68}:
  \bibinfo{pages}{63}. \bibinfo{doi}{\doi{10.1002/asna.18670680405}}.

\bibtype{Article}%
\bibitem[{Shao} and {Li}(2014)]{2014ApJ...796...37S}
\bibinfo{author}{{Shao} Y} and \bibinfo{author}{{Li} XD}
  (\bibinfo{year}{2014}).
\bibinfo{title}{{On the Formation of Be Stars through Binary Interaction}}.
\bibinfo{journal}{{\em \apj}} \bibinfo{volume}{796}, \bibinfo{eid}{37}.
  \bibinfo{doi}{\doi{10.1088/0004-637X/796/1/37}}.

\bibtype{Proceedings}%
\bibitem[Sigut and Jones(2016)]{2016ASPC..506.....S}
\bibinfo{editor}{Sigut TAA} and \bibinfo{editor}{Jones CE}, (Eds.)
  (\bibinfo{year}{2016}).
\bibinfo{title}{{Bright Emissaries: Be Stars as Messengers of Star-Disk
  Physics}}, \bibinfo{comment}{vol.} \bibinfo{volume}{506},
  \bibinfo{series}{Astronomical Society of the Pacific Conference Series}.

\bibtype{Article}%
\bibitem[{Slettebak}(1979)]{1979SSRv...23..541S}
\bibinfo{author}{{Slettebak} A} (\bibinfo{year}{1979}).
\bibinfo{title}{{The Be Stars}}.
\bibinfo{journal}{{\em \ssr}} \bibinfo{volume}{23}: \bibinfo{pages}{541--580}.
  \bibinfo{doi}{\doi{10.1007/BF00212356}}.

\bibtype{Article}%
\bibitem[{Slettebak}(1988)]{1988PASP..100..770S}
\bibinfo{author}{{Slettebak} A} (\bibinfo{year}{1988}).
\bibinfo{title}{{The Be Stars}}.
\bibinfo{journal}{{\em \pasp}} \bibinfo{volume}{100}: \bibinfo{pages}{770}.
  \bibinfo{doi}{\doi{10.1086/132234}}.

\bibtype{Article}%
\bibitem[{Smith} et al.(2016)]{2016AdSpR..58..782S}
\bibinfo{author}{{Smith} MA} et~al (\bibinfo{year}{2016}).
\bibinfo{title}{{The X-ray emission of the {\ensuremath{\gamma}} Cassiopeiae
  stars}}.
\bibinfo{journal}{{\em Advances in Space Research}} \bibinfo{volume}{58}:
  \bibinfo{pages}{782--808}. \bibinfo{doi}{\doi{10.1016/j.asr.2015.12.032}}.

\bibtype{Article}%
\bibitem[{Struve}(1931)]{1931ApJ....73...94S}
\bibinfo{author}{{Struve} O} (\bibinfo{year}{1931}).
\bibinfo{title}{{On the Origin of Bright Lines in Spectra of Stars of Class
  B}}.
\bibinfo{journal}{{\em \apj}} \bibinfo{volume}{73}: \bibinfo{pages}{94}.
  \bibinfo{doi}{\doi{10.1086/143298}}.

\bibtype{Article}%
\bibitem[{Suffak} et al.(2024)]{2024MNRAS.527.7515S}
\bibinfo{author}{{Suffak} MW} et~al (\bibinfo{year}{2024}).
\bibinfo{title}{{Disc tearing in a Be star: predicted 3D observations}}.
\bibinfo{journal}{{\em \mnras}} \bibinfo{volume}{527}:
  \bibinfo{pages}{7515--7522}. \bibinfo{doi}{\doi{10.1093/mnras/stad3659}}.

\bibtype{Book}%
\bibitem[{Underhill} and {Doazan}(1982)]{1982bsww.book.....U}
\bibinfo{author}{{Underhill} A} and \bibinfo{author}{{Doazan} V}
  (\bibinfo{year}{1982}).
\bibinfo{title}{B Stars with and without emission lines},
  \bibinfo{publisher}{NASA}.

\bibtype{Inproceedings}%
\bibitem[{Wade} et al.(2016)]{2016ASPC..506..207W}
\bibinfo{author}{{Wade} GA} et~al (\bibinfo{year}{2016}),
  \bibinfo{title}{{Magnetic Fields of Be Stars: Preliminary Results from a
  Hybrid Analysis of the MiMeS Sample}}, \bibinfo{editor}{{Sigut} TAA} and
  \bibinfo{editor}{{Jones} CE}, (Eds.), \bibinfo{booktitle}{Bright Emissaries:
  Be Stars as Messengers of Star-Disk Physics}, \bibinfo{series}{Astronomical
  Society of the Pacific Conference Series}, \bibinfo{volume}{506}, pp.
  \bibinfo{pages}{207}.

\bibtype{Article}%
\bibitem[{Wang} et al.(2021)]{2021AJ....161..248W}
\bibinfo{author}{{Wang} L} et~al (\bibinfo{year}{2021}).
\bibinfo{title}{{The Detection and Characterization of Be+sdO Binaries from
  HST/STIS FUV Spectroscopy}}.
\bibinfo{journal}{{\em \aj}} \bibinfo{volume}{161}, \bibinfo{eid}{248}.
  \bibinfo{doi}{\doi{10.3847/1538-3881/abf144}}.

\bibtype{Article}%
\bibitem[{Waters} and {Waelkens}(1998)]{1998ARA&A..36..233W}
\bibinfo{author}{{Waters} LBFM} and \bibinfo{author}{{Waelkens} C}
  (\bibinfo{year}{1998}).
\bibinfo{title}{{Herbig Ae/Be Stars}}.
\bibinfo{journal}{{\em \araa}} \bibinfo{volume}{36}: \bibinfo{pages}{233--266}.
  \bibinfo{doi}{\doi{10.1146/annurev.astro.36.1.233}}.

\bibtype{Article}%
\bibitem[{Zorec}(2023)]{2023Galax..11...54Z}
\bibinfo{author}{{Zorec} J} (\bibinfo{year}{2023}).
\bibinfo{title}{{BCD Spectrophotometry and Rotation of Active B-Type Stars:
  Theory and Observations}}.
\bibinfo{journal}{{\em Galaxies}} \bibinfo{volume}{11}, \bibinfo{eid}{54}.
  \bibinfo{doi}{\doi{10.3390/galaxies11020054}}.

\end{thebibliography*}

\end{document}